\documentclass[letterpaper,twoside,10pt]{article}
\usepackage{amssymb}
\usepackage{latexsym}
\usepackage{verbatim}
\usepackage{floatfig,epsfig}
\usepackage{calc}
\usepackage{fancyheadings}
\usepackage{pstricks,pst-node}
\usepackage{rotating,graphicx}
\newlength{\figurewidth}
\newcommand{\capitem}[1]{\caption{\textsf{#1}}}

\newcommand{\shortpage}{\enlargethispage{-\baselineskip}}
\newcommand{\papertitle}{%
Area-Constrained Planar Elastica%
}
\newcommand{\paperauthor}{%
G. Arreaga, R. Capovilla, C. Chryssomalakos and J. Guven%
}
\newcommand{\be}{\begin{equation}}
\newcommand{\ee}{\end{equation}}
\newcommand{\ble}[1]{\begin{equation} \label{#1}}
\newcommand{\bae}{\begin{eqnarray}}
\newcommand{\eae}{\end{eqnarray}}
\newcommand{\fle}[2]%
{\vspace{1.5ex}
\be
\label{#1}
\mbox{%
\setlength{\fboxsep}{3ex}%
\framebox{$\dss #2 $}}
\ee} 
\def\nn{\nonumber}
\newcommand{\ff}{\nn \\}
\newtheorem{exatitle}{Example}[section]
{\begin{exatitle}#1 \end{exatitle}}% 
{\hfill $\Box$ \\}
{\vspace{3mm} \noindent {\bf Aside:} \small}% 
{\hfill $\Diamond$ \\}
\def\th{\hat{t}}
\def\nh{\hat{n}}
\def\deps{\delta_\epsilon}
\def\derl{{d \over d \ell}}
\def\derk{{d \over d K}}

\def\fe{& = &}

\newcommand{\dss}{\displaystyle}
\def\fedss{& = & \dss}
\newcommand{\aqqadss}{& \qquad & \dss}
\newcommand{\calO}{\mathcal{O}}
\newcommand{\calI}{\mathcal{I}}
\def\eg{\hbox{\it e.g.}}

\def\ie{\hbox{\it i.e.}}
\newcommand{\K}{\mathcal{K}}
\begin{document}
\initfloatingfigs
%%%%%%%%%%%%%%%%%%%%%%%%%%%%%%%%%%%%%%%%%%%%%%%%%%%%%%%%%%%%%%
%%%%%%%%%%%%%%%%%%%%%%%%%%%%%%%%%%%%%%%%%%%%%%%%%%%%%%%%%%%%%%
%%%%%%%%%%%%%%%%%%%%%%%%%%%%%%%%%%%%%%%%%%%%%%%%%%%%%%%%%%%%%%
% Titlepage
%%%%%%%%%%%%%%%%%%%%%%%%%%%%%%%%%%%%%%%%%%%%%%%%%%%%%%%%%%%%%%
\begin{titlepage}
\vspace*{-1cm}
\begin{flushright}
\textsf{CIEA-GR-01-02}
\\
\textsf{ICN-UNAM-01/03}
\\
\mbox{}
\\
\textsf{July 5, 2001}
\\[3cm]
\end{flushright}
%%%%%%%%%%%%%%%%%%%%%%%%%%%%%%%%%%%%%%%%%%%%%%%%%%%%%%%%%%%%%%
%%%%%%%%%%%%%%%%%%%%%%%%%%%%%%%%%%%%%%%%%%%%%%%%%%%%%%%%%%%%%%
%%% TITLE, AUTHORS
%%%%%%%%%%%%%%%%%%%%%%%%%%%%%%%%%%%%%%%%%%%%%%%%%%%%%%%%%%%%%%
%%%%%%%%%%%%%%%%%%%%%%%%%%%%%%%%%%%%%%%%%%%%%%%%%%%%%%%%%%%%%%
%\begin{center}
%%%%%%%%%%%%%%%%%%%%%%%%%%%%%%%%%%%%%%%%%%%%%%%%%%%%%%%%%%%%%%
\renewcommand{\thefootnote}{\fnsymbol{footnote}}
\begin{LARGE}
%\bfseries{\sffamily Equilibrium Configurations of a Rigid Loop}
%\bfseries{\sffamily Equilibria of Elastica Hypoarealis}
\bfseries{\sffamily \papertitle}
\end{LARGE}

\noindent \rule{\textwidth}{.6mm}

\vspace*{1.6cm}

\noindent \begin{large}%
\textsf{\bfseries%
Guillermo Arreaga, Riccardo Capovilla,
}
\end{large}

%\vspace*{.1cm}

\phantom{XX}
\begin{minipage}{.8\textwidth}
\begin{it}
\noindent Departamento de F\'\i sica\\
CINVESTAV IPN\\
Apdo. Postal 14740\\
07000 M\'exico, D.F., MEXICO\\
\end{it}
\texttt{garreaga@fis.cinvestav.mx, capo@fis.cinvestav.mx
\phantom{X}}
\end{minipage}

\vspace*{1cm}
\noindent
\begin{large}%
\textsf{\bfseries%
Chryssomalis Chryssomalakos and Jemal Guven
}
\end{large}

%\vspace*{.1cm}

\phantom{XX}
\begin{minipage}{.8\textwidth}
\begin{it}
\noindent Instituto de Ciencias Nucleares \\
Universidad Nacional Aut\'onoma de M\'exico\\
Apdo. Postal 70-543, 04510 M\'exico, D.F., MEXICO \\
\end{it}
\texttt{chryss@nuclecu.unam.mx, jemal@nuclecu.unam.mx
\phantom{X}}
\end{minipage}
\\

\vspace*{3cm}
%%%%%%%%%%%%%%%%%%%%%%%%%%%%%%%%%%%%%%%%%%%%%%%%%%%%%%%%%%%%%%
%%% ABSTRACT
%%%%%%%%%%%%%%%%%%%%%%%%%%%%%%%%%%%%%%%%%%%%%%%%%%%%%%%%%%%%%%
\noindent
\textsc{\large Abstract: }
We determine the equilibria of a rigid 
loop in the plane, subject to the constraints of fixed length 
and {\em fixed enclosed area}.
Rigidity is characterized by an energy functional quadratic in 
the curvature of the loop.
We find that the area constraint gives rise to equilibria with 
remarkable geometrical properties:   
not only can the Euler-Lagrange equation be integrated to provide 
a quadrature for the curvature but, in addition,
the embedding itself can be expressed as a local function of the  
curvature. The configuration space is shown to
be essentially one-dimensional, with surprisingly rich structure.
Distinct branches of integer-indexed equilibria
exhibit self-intersections and bifurcations --- a gallery of plots 
is provided to highlight these findings.
Perturbations connecting equilibria are shown to satisfy 
a first order ODE which is readily solved. 
We also obtain analytical expressions 
for the energy as a function of the area
in some limiting regimes. 

\end{titlepage}
\setcounter{footnote}{1}
\renewcommand{\thefootnote}{\arabic{footnote}}
\setcounter{page}{2}
%%%%%%%%%%%%%%%%%%%%%%%%%%%%%%%%%%%%%%%%%%%%%%%%%%%%%%%%%%%%%%%%%
%%%%%%%%%%%%%%%%%%%%%%%%%%%%%%%%%%%%%%%%%%%%%%%%%%%%%%%%%%%%%%%%%
%%%%%%%%%%%%%%%%%%%%%%%%%%%%%%%%%%%%%%%%%%%%%%%%%%%%%%%%%%%%%%%%%
\noindent \rule{\textwidth}{.5mm}

\tableofcontents

\noindent \rule{\textwidth}{.5mm}
\section{Introduction}
\label{Intro}
%%%%%%%%%%%%%%%%%%%%%%%%%%%%%%%%%%%%%%%%%%%%%%%%%%%%%%%%%%%%%%%%
%%%%%%%%%%%%%%%%%%%%%%%%%%%%%%%%%%%%%%%%%%%%%%%%%%%%%%%%%%%%%%%%%
Our object of study in this paper is a deflated rigid loop 
({\em elastica hypoarealis}). 
To explain this, consider a closed loop in a plane. 
The loop is made of some
elastic material and, if taken thin enough, the contribution of its 
longitudinal deformations to its elastic energy can be ignored.
This gives rise to a hamiltonian density proportional to the
square of its curvature --- hence {\em elastica}. 
It is clear that fixing only the length $L$ of the loop will not 
lead to any surprises, the only possible equilibrium
being a circle. What will happen though if we, in addition,
fix its area $A$ to be less than that of the circle? (hence {\em
hypoarealis}).  

This particular hamiltonian (without the constraints) has a 
distinguished history,
dating back at least to the Bernoullis. It makes a reappearance in
Euler's inspired analysis of a
mundane but subtle problem in mechanical engineering:
the buckling of a loaded beam \cite{Tru:83, Gia.Hil:96}.
More recently, variants of this problem have attracted the 
attention of mathematicians. In particular, 
the problem of determining the curves of constant length 
which minimize the bending energy, on surfaces 
of constant curvature, is considered in
Refs.~\cite{Gri:83,Bry.Gri:86,Lan.Sin:84}, while interconnections
with knot theory are explored in~\cite{Ive.Sin:97}. 
Adding the constraint on 
the enclosed area, as we propose here, provides a 
particularly fruitful generalization of this work. 

Hamiltonians depending on extrinsic 
curvature have also been studied in 
statistical physics~\cite{Amb.Dur.Jon:97,Saf:94}. 
For space curves, 
an effective hamiltonian which depends on twist as well as bending
provides a phenomenological description of stiff polymers, 
and in particular DNA (see {\it e.g.}
Refs.~\cite{Mar.Sig:95,Bou.Mez:98}). 
Moving up one dimension, the leading term in the Helfrich 
hamiltonian, which describes 
the equilibrium configurations of lipid membranes, is proportional to
the square of the traced extrinsic curvature $K$ integrated
over the surface, namely, the conformally invariant 
Willmore functional~\cite{Wil:82,Hel:73,Pel:94,Sei:97}.
In the variant of the Helfrich model proposed by Svetina and Zeks,
the so-called bilayer couple model, 
constraints are placed on the surface, such as constant area, constant
enclosed volume, and constant integrated mean curvature 
(the latter breaking  the $K \to - K$ 
symmetry of the 
problem)~\cite{Sve.Zek:89}.\footnote{ 
We note that there is no genuine
one-dimensional
analog of its refinement, the so-called
area difference model \cite{Sei:97}.}  However, the equations 
which determine
the equilibrium are highly non-trivial higher-order PDEs.
One motivation for introducing our variant of the {\em
elastica}, as a planar analogue of the above, is its potential
as a toy model for understanding these higher dimensional membranes.
At the simplest level, if a closed membrane possesses a symmetry along
the $z$-axis, its profile at any $z$ will be a fixed loop.
The dimensionally reduced Helfrich hamiltonian is then the 
bending energy of 
this loop. It is possible to examine the loop
hamiltonian exactly.
While a lot is known about axially symmetrical closed
configurations~\cite{Deu.Hel:76},
there is very little non-perturbative knowledge of
the equilibria which exist when axial symmetry is broken:
the loop configurations provide an analytic point of
entry.
 
An additional application of the planar problem was pointed out 
recently by Willmore: 
any closed loop can be exploited to generate an  axially symmetric 
toroidal geometry. The equilibria of the loop 
hamiltonian on a surface of constant negative curvature can be 
mapped into
equilibria of the Helfrich functional \cite{Wil:00}. 

The feature of our model we will exploit
is its integrability, very well disguised when
the problem is cast with respect to the variables embedding the 
loop in
the plane. Indeed, the determination of the curvature at equilibrium
can be reduced to the study of the motion of a particle in a
one-dimensional quartic potential.
The curvature can be solved as a quadrature in terms of
elliptic integrals. 
Remarkably, we discover that the problem possesses a second, 
far less-obvious, level
of integrability: the embedding can be expressed as a 
local function of the  
curvature without any need of the two integrations one 
would have expected. As a consequence, 
the equilibrium configuration 
is given by a geometrical construction of 
pythagorean simplicity.

The condition of closure of the loop 
results in a discrete spectrum for a given
constrained perimeter and area.  We show that 
scaling can be exploited to fix one of these, the perimeter say.
One branch of the spectrum then consists of configurations with 
an $n$-fold symmetry ($n=2,3,\ldots$) which inflate into   
a circle as the enclosed area is increased. 
If the area is evacuated these configurations will 
eventually self-intersect, tending to a 
limiting configuration  of an inverted (\ie, negative area) circle,   
winding $n-1$ times and 
decorated with $n$ infinitesimally 
small circles. The latter dominate, in this limit,  
the energy and give rise to a pole in the
energy versus area diagram, the residue of which 
we evaluate exactly for each $n$. 

The remaining branches of the spectrum 
correspond to rather complicated self-intersecting 
pairs which 
bifurcate from a limiting configuration. While self-intersecting 
configurations are 
undoubtedly of interest mathematically, they are not of primary
relevance to the physical problem we pose --- we determine sufficient 
conditions for avoiding them (nevertheless, we describe a
physical situation where they may arise).

In general, one would expect equilibria connecting perturbations to be 
governed by a fourth order
differential equation. We show, however, that 
this equation can be reduced 
(by three orders) to
a first order one. The latter
is singular at the circular configuration,
where bifurcation occurs.   
Analytical approximations for the energy as a function of 
the area are derived, 
in the limit of sinusoidal perturbations of the circle of frequency
$n$. 
Combined with the pole mentioned above, these expressions
provide a reliable sketch 
of the energy -- area relationship for each $n$.

The paper is organized as follows:
Sect.~\ref{FC} introduces the model and the shape equation that
determines equilibria. We show how it can be reduced to
the motion of a fictitious classical particle in a 
quartic potential. We address the behaviour under scaling of
the shape equation and we show that the problem has essentially
one free parameter. Moreover we analyze the loop statics; this
provides a physical interpretation for the shape equation itself. 
In Sect.~\ref{Config} we study the configuration space for this model.
In particular, we describe the angle $\Theta_0$ by which the
normal to the loop gets rotated in a full oscillation
of the fictitious particle in the potential. Equilibria
connecting perturbations is the subject of Sect.~\ref{PA}. These
are determined by a complicated fourth order differential
equation, which
is used to obtain the purely geometrical construction 
mentioned above. In turn, the latter permits the reduction of the order
of the original equation from four to one, as well as the 
derivation of a 
sufficient condition for non-self-intersections. Finally,
we obtain analytical expressions for the energy as a function of
the  area, in some limiting regimes, which allow for a reliable
sketch of its behaviour. In the appendices we collect various 
expressions, useful in the calculation of variations, we derive 
a recursion relation for the average of the powers of $K$
and comment on an interesting Legendre transform in the space of
parameters of the model.
%%%%%%%%%%%%%%%%%%%%%%%%%%%%%%%%%%%%%%%%%%%%%%%%%%%%%%%%%%%%%%%%%
%%%%%%%%%%%%%%%%%%%%%%%%%%%%%%%%%%%%%%%%%%%%%%%%%%%%%%%%%%%%%%%%%
\section{First Considerations} 
\label{FC}
%%%%%%%%%%%%%%%%%%%%%%%%%%%%%%%%%%%%%%%%%%%%%%%%%%%%%%%%%%%%%%%%
%%%%%%%%%%%%%%%%%%%%%%%%%%%%%%%%%%%%%%%%%%%%%%%%%%%%%%%%%%%%%%%%%
%%%%%%%%%%%%%%%%%%%%%%%%%%%%%%%%%%%%%%%%%%%%%%%%%%%%%%%%%%%%%%%%%
\subsection{Energy functional} 
\label{ef}
%%%%%%%%%%%%%%%%%%%%%%%%%%%%%%%%%%%%%%%%%%%%%%%%%%%%%%%%%%%%%%%%
A closed loop, parametrized by $s \in [0,1]$, is described by 
the embedding in the plane,
\[
\vec x = \vec X(s)
\, .
\]
The arc length $\ell$ along the loop is given by
\begin{equation}
\ell(s) = \int_0^s ds' \,\left({d \vec X\over ds'}\cdot
{d \vec X\over ds'}\right)^{1/2}
\,.
\end{equation}
The most general expression for the energy of the
configuration $X(s)$
which {\em i}) does not depend on the parametrization,
{\em ii}) involves no higher than two derivatives of $\vec X$ and
{\em iii}) is
quadratic in these derivatives, is given by
\begin{equation}
\label{Fzdef}
F[\vec X]=
\alpha \int  d\ell\, K^2\,,
\end{equation}
where $K$ is the geodesic curvature, equal to the inverse
of the radius of curvature at each point of the loop.
We will take henceforth $\alpha$ to be equal to 1 ---
$\alpha$ is dimensionful, unlike its two-dimensional
analogue, and this means we are measuring length in units
of $\alpha$.
Let $\th$ be the unit tangent to the loop (transversed
counterclockwise) and $\nh$ its outwards
normal. Then we have the Frenet-Serret equations for
a plane curve\footnote{Notice that the opposite sign convention
for $K$ is quite common in the literature.} (the prime denotes a 
derivative with respect to arc length $\ell$),
\begin{equation}
\nh'=K\th 
\, , 
\qquad \qquad
\qquad \qquad
\th'=-K\nh
\, .
\end{equation}
In terms of the angle $\Theta$ that $\nh$ makes with,
say, the $x$-axis, $K=\Theta'$. 

To implement the constraints of fixed length and
enclosed area,
we introduce the constrained functional:
\begin{equation}
\label{eq:k1}
F_c[X]=
F[X]
+ \mu \left(\int d\ell - L \right)
+ \sigma \left(\int_{\rm int}d^2x -A \right)\,.
\end{equation}
$\mu$ appears as a lagrange
multiplier enforcing the constraint
fixing the length of the loop to some value, $L$.
In the same way, $-\sigma$ is associated
with the constraint fixing the enclosed area to the value
$A$. We note that, in general,
\be
\label{musigma}
\mu  = -{\partial F_c(L, A)\over \partial L}
\, , \qquad \qquad
\sigma  = -{\partial F_c(L, A)\over \partial A}
\,.
\ee
In particular, if the area constraint is relaxed so that
$\sigma=0$, then
$\partial F_c(L, A)/ \partial A= 0$. Looking at
(\ref{musigma}), one might be led to
identify $\mu$, $\sigma$ with the tension and differential
pressure on the loop but this is only half true --- we discuss the
physical meaning of these parameters in Sec.{} \ref{oft}.
We have not included, in our energy functional, a term proportional 
to the integral of $K$. This is because
such a term is simply the total angle by which $\nh$ gets
rotated in transversing the loop once, equal to $2 \pi$
for a non-self-intersecting loop.

We emphasize that the bending energy
is not scale invariant
(notice that neither is the dimensionally reduced Helfrich
hamiltonian). An unconstrained circular loop will expand
without limit.
The natural scale invariant
expression, $\int d\ell\, | K |$,
though not in itself a topological invariant, does have a
vanishing
Euler-Lagrange derivative almost everywhere. Taken as an
energy functional, its
constrained configurations will be arcs of a circle
(solutions of
$\mu K= \sigma$ ) joined together at curvature
discontinuities
in such a way as to mimimize the corresponding
constrained energy
function. This model will be considered elsewhere.
%%%%%%%%%%%%%%%%%%%%%%%%%%%%%%%%%%%%%%%%%%%%%%%%%%%%%%%%%%%%%%%%%
\subsection{Shape equation} 
\label{eom}
%%%%%%%%%%%%%%%%%%%%%%%%%%%%%%%%%%%%%%%%%%%%%%%%%%%%%%%%%%%%%%%%%
The Euler-Lagrange equations follow from extremizing $F_c[X]$,
\begin{equation}
{\delta F_c \over \delta X^\mu}=0\,.
\end{equation}
The reparametrization invariance of $F_c[X]$
implies that the tangential projection of these equations
is vacuous. 
The normal projection gives 
\fle{eq:D1}{%
2  K'' + K^3 - \mu K -  \sigma = 0 
}%
(see the appendix for some relevant formulae).
We note that with the identification, $\mu\to\mu +
K_0^2$,
Eq.~(\ref{eq:D1}) coincides with Eq.~(7) of 
\cite{Hel.OuY:87} (see also
\cite{Hel.OuY:89}). Is is worth pointing out that a loop on a
surface of constant Gaussian curvature $G$ satisfies the above
equation with $\mu \rightarrow \mu - 2G$.
 
A plane curve is determined, up to rigid motions, by
its curvature \cite{Car:76}. 
As is well known,
our one-dimensional system is completely integrable. 
This
is obvious
because we have cast derivatives with respect to arclength.
With respect to
an arbitrary parametrization of the loop, first
derivative terms
associated with the one-dimensional Laplacian appear
which conceal this fact. 
Writing~(\ref{eq:D1}) in the form 
\be
\label{DEKalt}
K'' = - \derk V(K)
\, , 
\qquad
\qquad
V(K) \equiv \frac{1}{8} K^4 - \frac{\mu}{4} K^2 -\frac{\sigma}{2} K
\, ,
\ee
we  map our problem onto the problem of
determining the motion of a fictitious particle in a quartic potential,
$K$
being the displacement of the particle and $\ell$ playing the role 
of time. The total energy $E$ of the particle is conserved,
\be
\label{eq:D2}
{dE \over d \ell} = 0
\, , \qquad \qquad
E \equiv {1 \over 2} {K'}^2 + V(K) 
\, ,
\ee
a fact that permits the expression of $K'$ in terms of $K$ and
hence, of the arclength along the loop as an integral over $K$ (via
$d \ell = dK/K'$)
\be
\label{ellintK}
\ell = \int \frac{dK}{\sqrt{2(E-V(K))}}
\, .
\ee
We emphasize the difference between the energy $E$ of the fictitious 
particle, on the one hand,  and 
the bending energy $F$ of the loop, on the other.
In particular, the configuration
of least `energy' in the analogue is not the minimum of the
hamiltonian.

The motion of the particle in the potential $V$ is periodic. 
%A moment's thought reveals that 
Closed loop configurations consist
of an integer number of identical segments, each one
corresponding to one full oscillation of the particle, and hence 
made up itself of two symmetric halves. Then, the
condition of closure of the loop can be expressed as
\be
\label{condclos}
\Theta_0 = \frac{2 \pi}{n}
\, , \qquad
n=2,3,\dots
\ee
where $\Theta_0$ is the angle by which the normal
$\nh$ gets rotated in one
full oscillation of the particle, given by
\be
\label{Thetaint}
\Theta_0 = 2 \int_{K_{min}}^{K_{max}} \frac{KdK}{\sqrt{2(E-V(K))}}
\ee
($K_{min}$, $K_{max}$ denote the turning points).
The equilibrium configurations have $n$-fold symmetry and
a well-defined center. 
The value $n=1$ is omitted in~(\ref{condclos}) because it is
special -- see below.  We examine in detail the
resulting configurations in Sect.~\ref{Config}.
%%%%%%%%%%%%%%%%%%%%%%%%%%%%%%%%%%%%%%%%%%%%%%%%%%%%%%%%%%%%%%%%%
\subsection{Scaling}
\label{Scaling}
%%%%%%%%%%%%%%%%%%%%%%%%%%%%%%%%%%%%%%%%%%%%%%%%%%%%%%%%%%%%%%%%
Let us examine the behavior of Eq.~(\ref{eq:D1}) under
scaling of the position vector, $\vec X\to \lambda \vec X$. 
We find
\begin{equation}
\label{eq:kscale}
\ell \to \lambda \ell
\, , \qquad \qquad
{d \over d \ell} \to {1 \over \lambda} {d \over d \ell}
\, , \qquad \qquad
K \to {1\over \lambda} \, K\,.
\end{equation}
Let $\vec X(\ell)$ correspond to some given solution of
Eq.~(\ref{eq:D1}), with parameter 
values $(\sigma_1,\mu_1)$.
Then, a scaled solution with the same
shape, $\lambda \vec X(\lambda\ell)$
and with ($L_1$, $A_1$, $F_1$, $E_1$) $\to$ ($\lambda L_1$,
$\lambda^2 A_1$, $\lambda^{-1} F_1$, $\lambda^{-4} E_1$), 
is obtained by rescaling the multipliers as follows:
\begin{equation}
\label{eq:Pscale}
\mu_1 \to {1\over \lambda^2} \mu_1 \equiv \mu_\lambda
\, , 
\qquad
\qquad
\qquad
\sigma_1 \to {1\over \lambda^3} \sigma_1 \equiv \sigma_\lambda
\,.
\end{equation}
Eliminating $\lambda$ we find the orbits of scaling in the
$(\sigma,\mu)$-plane
\be
\label{scaleorb}
\mu_\lambda = \frac{\mu_1}{\sigma_1^{2/3}} \, 
\sigma_\lambda^{2/3}
\, .
\ee
Furthermore, an inversion in the origin,
$X^\mu \to - X^\mu$ can be identified with a rescaling by
$\lambda= -1$, which maps a solution with a given $\sigma$
into one with $-\sigma$.
It follows that, for the purpose of identifying distinct
configurations at least, one can set, \eg \ $\sigma=1$ and scan the
essentially one-dimensional configuration space by varying
$\mu$. For each value of $\mu$, which fixes the form of the
potential $V(K)$, one has still to vary the energy
$E$ of the particle in the well so as to satisfy~(\ref{condclos}),
resulting, in general, in an infinite discrete spectrum ``above''
the point $(1,\mu)$ in the $(\sigma,\mu)$-plane. 
%%%%%%%%%%%%%%%%%%%%%%%% FIGURE
\begin{figure}
%\framebox{%
%%%%%%%%%%%%%%%% PSPICTURE
\begin{pspicture}(0,0)(.16\textwidth,.3\textwidth)
%%%%%%%%% Units
\psset{unit=.1\textwidth}
\setlength{\unitlength}{.1\textwidth}
%\psgrid[subgriddiv=10,griddots=10,gridlabels=10pt]
%%%%%%%%% Puts (labels)
\put(.1,.35){\makebox[0mm][l]{$\sigma=1$}}
\put(.1,.13){\makebox[0mm][l]{$\bar{\mu}=0.5$}}
\put(1.95,2.8){\makebox[0mm][r]{$\bar{V}$}}
\put(3.8,1.4){\makebox[0mm][l]{$\bar{K}$}}
\put(3.62,1.75){\makebox[0mm][l]{$n \! = \! 2$}}
\put(3.65,1.93){\makebox[0mm][l]{$n \! = \! 3$}}
\put(3.67,2.08){\makebox[0mm][l]{$n \! = \! 4$}}
\put(3.69,2.26){\makebox[0mm][l]{$n \! = \! 5$}}
%%%%%%%%% Axes
\psline[linewidth=.3mm]{->}%
(.4,1.5)(3.75,1.5)
\psline[linewidth=.3mm]{->}%
(1.71,.3)(1.71,2.85)
%%%%%%%%% Lines
\psline[linewidth=.2mm]{*-*}%
(1.25,1.84)(3.54,1.84)
\psline[linewidth=.2mm]{*-*}%
(1,2)(3.57,2)
\psline[linewidth=.2mm]{*-*}%
(.85,2.12)(3.59,2.12)
\psline[linewidth=.2mm]{*-*}%
(.72,2.26)(3.61,2.26)
%\rput(3,2.3){\rnode{A}{}}
%\rput(4.3,1.83){\rnode{B}{}}
%\ncarc{->}{A}{B}
%%%%%%%%% Included graphic
%\framebox{
\raisebox{\totalheight}{%
\rotatebox{270}{%
%\framebox{%
\includegraphics[width=.3\textwidth]{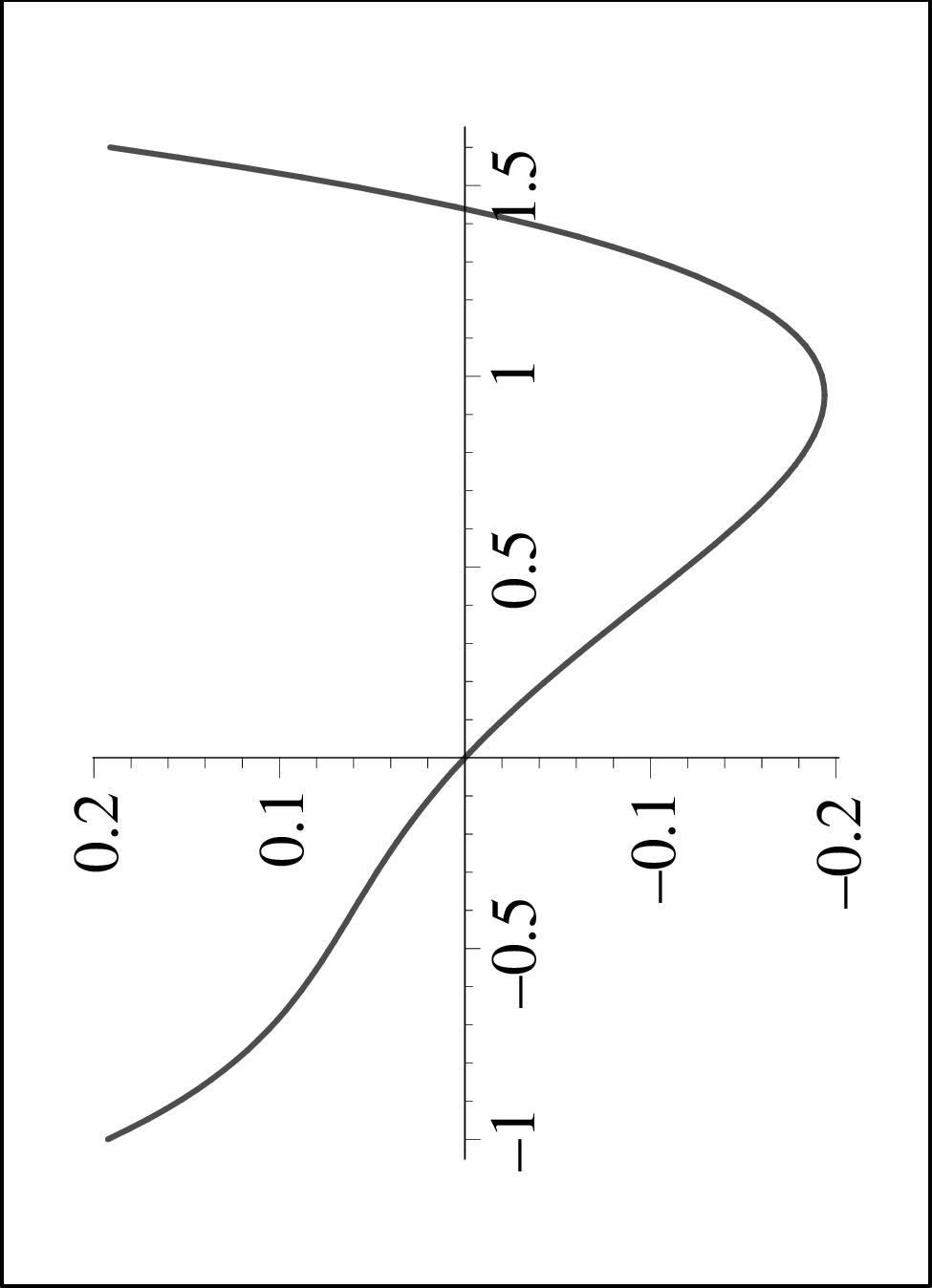}%
}%
%}%
%}
}
\end{pspicture}
%%%%%%%%%%%%%%%% PSPICTURE
%}
\hfill
%%%%%%%%%%%%%%%% PSPICTURE
\begin{pspicture}(0\textwidth,0\textwidth)(.4\textwidth,.3\textwidth)
%%%%%%%%% Units
\psset{unit=.1\textwidth}
\setlength{\unitlength}{.1\textwidth}
%\psgrid[subgriddiv=10,griddots=10,gridlabels=10pt]
%%%%%%%%% Axes
\psline[linewidth=.3mm]{->}%
(.4,1.5)(3.85,1.5)
\psline[linewidth=.3mm]{->}%
(1.563,.3)(1.563,2.75)
%%%%%%%%% Lines
%%%%%%%%% Puts
\put(.1,.35){\makebox[0mm][l]{$\sigma=1$}}
\put(.1,.13){\makebox[0mm][l]{$\bar{\mu}=0.5$}}
\put(1.5,2.8){\makebox[0mm][r]{$\bar{K'}$}}
\put(3.8,1.27){\makebox[0mm][l]{$\bar{K}$}}
\put(2.3,2.23){\makebox[0mm][l]{$n \! = \! 2$}}
\put(2.13,2.55){\makebox[0mm][r]{$n \! = \! 5$}}
%%%%%%%%% Included graphic
\raisebox{\totalheight}{%
\rotatebox{270}{%
\includegraphics[width=.3\textwidth]{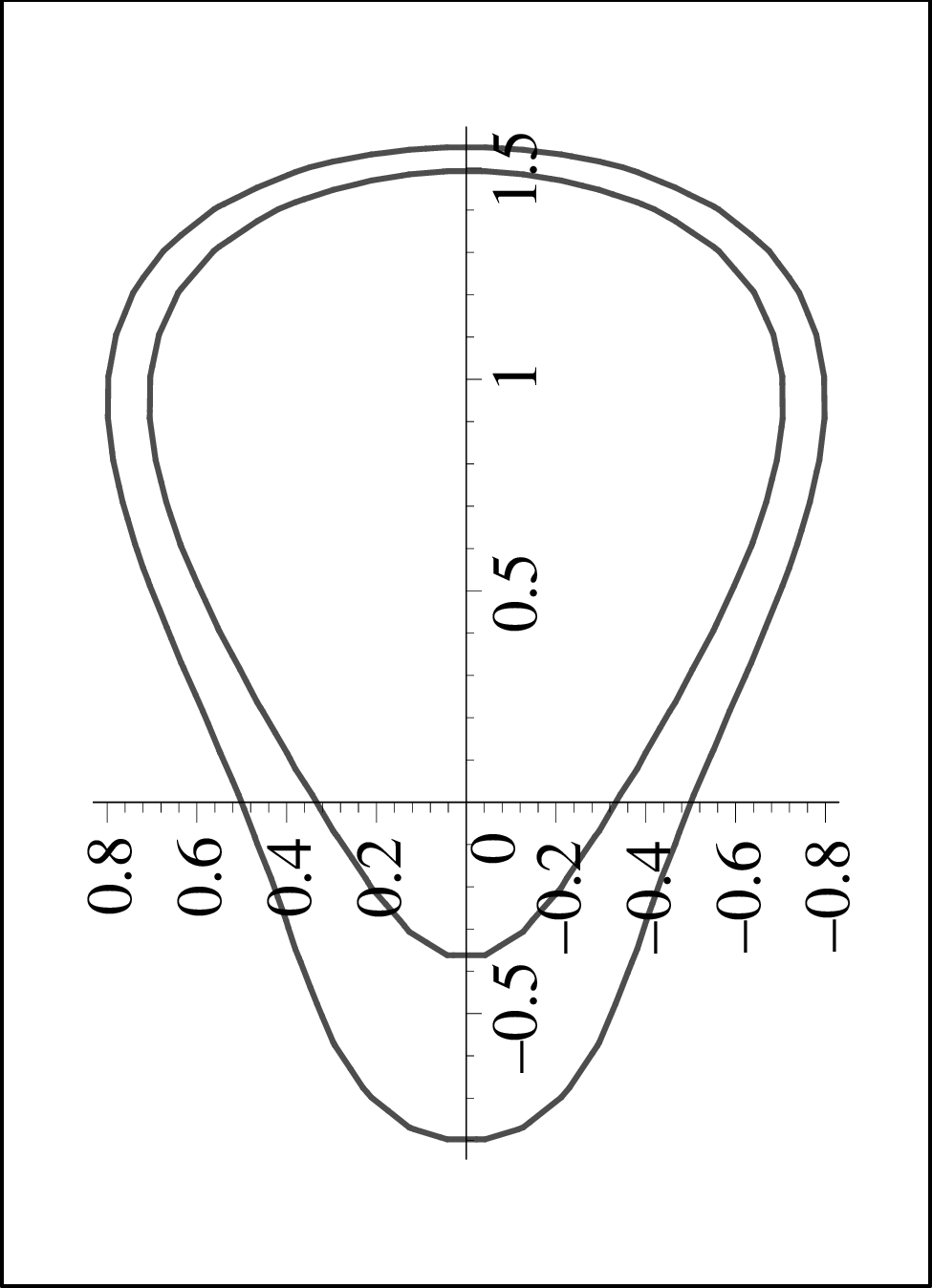}%
}%
}
\end{pspicture}
%%%%%%%%%%%%%%%% PSPICTURE
%%%%%%%%%%%%%%%% CAPTION
\capitem{%
On the left: the potential $\bar{V}(\bar{K})$, for the values of
$\bar{\mu}$, $\sigma$ shown. Barred 
quantities, in this and subsequent figures, are
measured in units of $\mu_0^{-q/2}$, $q$ being their length
dimension. Also shown, with horizontal lines, are the energies
$\bar{E}^{(n)}$ 
that give rise to the $n=2,3,4,5$ closed configurations, equal to
0.253, 0.278, 0.298, 0.320 respectively (in units of $\mu_0^2$ and
{\em measured from the bottom of the well} --- compare
with the $\bar{\mu}=0.5$ curve in Fig.{}~\ref{Theta0Rpm}). 
On the right: the closed trajectories in the phase plane
$\bar{K}-\bar{K'}$ of the
ficticious particle, corresponding to oscillations in the
potential $V$ with energies $\bar{E}^{(2)}$, $\bar{E}^{(5)}$ ---
the $n=3,4$ trajectories lie in between the two shown.
}
\label{phase}
\end{figure}
%%%%%%%%%%%%%%%%%%%%%%%% FIGURE
%%%%%%%%%%%%%%%%%%%%%%%%%%%%%%%%%%%%%%%%%%%%%%%%%%%%%%%%%%%%%%%%%
\subsection{Windings} 
\label{Wind}
%%%%%%%%%%%%%%%%%%%%%%%%%%%%%%%%%%%%%%%%%%%%%%%%%%%%%%%%%%%%%%%%%
The closed configurations we wish to classify possess two
topological invariants, both given as the winding numbers of maps
$S^1 \to S^1$. The first such map is the normal map of the loop,
associating with each value of the parameter $s$ (which ranges
over the first $S^1$ above) the corresponding value of $\nh$ (which
ranges over the unit circle, the second $S^1$ above). For each
rotation of $s$, $\nh$ will generally rotate $m$ times, with
$m$ an integer. 
Configurations with distinct $m$'s are topologically isolated,
the physical implementation being the infinite energy barrier 
associated with the move
%%%%%%%%%%%%% PS PICTURE
\begin{pspicture}(0,0)(5.8,.3)
%\pssetlength{\linewidth}{.1mm}
\psset{xunit=1cm,yunit=.4cm}
\pscurve[linewidth=.2mm]{<->}%
(0,0)(.7,.6)(.5,.8)(.3,.6)(1,0)
\pscurve[linewidth=.2mm]{->}%
(1.3,.3)(1.6,.3)(1.8,.3)
\pscurve[linewidth=.2mm]{<->}%
(2.1,0)(2.65,.75)(2.6,.8)(2.55,.75)(3.1,0)
\pscurve[linewidth=.2mm]{->}%
(3.4,.3)(3.7,.3)(3.9,.3)
\pscurve[linewidth=.2mm]{<->}%
(4.2,0)(4.8,.3)(5,.7)(5.2,.3)(5.8,0)
\end{pspicture}
%%%%%%%%%%%%% PS PICTURE
. Non-self-intersecting loops correspond to normal maps with 
index $m=\pm 1$ (the converse is not true).

The second map is from the $s$-circle to the  closed orbit 
($\sim S^1$) traced by the phase point in the plane $(K,K')$. 
For a configuration
corresponding to $n$ complete oscillations of the particle, 
the phase point goes around the closed curve $\frac{1}{2} {K'}^2
+ V(K)=E$ $n$ times, Fig.{}~\ref{phase}.
There is a finite energy barrier separating configurations 
with distinct $n$'s which prevents transitions between them.  
%%%%%%%%%%%%%%%%%%%%%%%%%%%%%%%%%%%%%%%%%%%%%%%%%%%%%%%%%%%%%%%%%
\subsection{Forces and torques} 
\label{oft}
%%%%%%%%%%%%%%%%%%%%%%%%%%%%%%%%%%%%%%%%%%%%%%%%%%%%%%%%%%%%%%%%%
What can we learn about the loop configurations by looking at the
equilibrium of forces and torques on an infinitesimal loop
segment? First we ask, why would a rigid loop want to have area less
than the maximal allowed by its length? One setup that supplies 
an answer is to imagine that our loop is actually the
cross-section of an infinite cylinder, the interior {\em and} 
exterior of
which are filled with an incompressible fluid. We start by
filling the cylinder to its maximum capacity, this gives a
circular loop. We then take out some fluid from the interior 
while making sure that the external pressure is large enough so as
to prevent the formation of bubbles in the interior.
Notice that, because of its
rigidity, the cylinder takes up some part of the exterior
pressure and only transmits to its interior a fraction of it, it's
the differential pressure that crumbles the cylinder walls. The 
second of~(\ref{musigma}) then points to the identification of
$-\sigma$ with this differential pressure, taking into account
that the latter points inwards. One has to be careful in applying
the same argument to $\mu$. The first of (\ref{musigma}) seems to
suggest that $\mu$ is the tension of the loop but this would
presuppose the possibility of tangential deformations --- there
is nothing in our energy functional that tells us how much energy
these cost and, indeed, we have already assumed that the loop
{\em cannot} be stretched or compressed tangentially. The
derivative in the first of~(\ref{musigma}) is computed by
comparing distinct loops with infinitesimally differing  lengths, 
it does not refer
to the deformation of a single loop.  With this in mind, we now
turn to loop statics.

We denote by $\vec T(\ell_0)$  the total force from the
segment of the loop with $\ell < \ell_0$ to the one with $\ell >
\ell_0$ and by $\tau(\ell_0)$ the corresponding torque. The
latter is equal to $-2K$ --- this follows from our normalization of
the bending energy $F$. Then,
balancing the torques on a segment extending from $\ell$ to $\ell +
d \ell$ we get
\ble{torquebal}
\tau(\ell) - \tau(\ell + d \ell) + T_n(\ell + d \ell) d\ell =0
\qquad \Rightarrow \qquad
\tau' - T_n = 0
\, ,
\ee
where $T_n$ is the normal force, the reference point was taken at $\ell$ and the sign
conventions are shown in Fig.{}~\ref{torqfig}. 
%%%%%%%%%%%%%%%%%%%%%%%% FIGURE
\begin{figure}
\resizebox{\textwidth}{.28\textwidth}{%
\setlength{\fboxsep}{4mm}
\framebox{%
%%%%%%%%%%%%%%%%%% PS PICTURE: Torque balancing
\begin{pspicture}(0,0)(.33\textwidth,.33\textwidth)
%%%%---- unit lengths for Latex, pstricks
\psset{unit=.1\textwidth}
\setlength{\unitlength}{.1\textwidth}
%\psgrid[subgriddiv=10,griddots=10,gridlabels=10pt]
%%%%---- Main curve
\pscurve[linewidth=.3mm,showpoints=false]{-}%
(0.6,0.6)(.93,1.5)(1,2)(.9,2.5)(.6,2.85)(0,3)
%%%%---- Lines
\psline[linewidth=.3mm]{->}%
(1,2)(1,3)
\psline[linewidth=.3mm]{->}%
(1,2)(2,2)
\psline[linestyle=dashed]{-}%
(.9,2.5)(2.5,2.5)
%\psline{->}%
%(1,2)(2.6,2)
\psline{->}%
(.9,2.5)(2.27,3.2)
%%%%---- Positive direction for torques
\psarc{->}(1.4,1.2){.2}{-180}{90}
%%%%---- Actual directions for torques, angle
\psarcn{->}(1,2){.1}{90}{-180}
\psarc{->}(.9,2.5){.1}{0}{270}
\psarc{->}(.9,2.5){.5}{0}{26}
%%%%---- Labels, dots
\put(1.33,1.16){$+$}
\put(1.1,3){$\th$}
\put(1.95,1.8){$\nh$}
\psdots(1,2)(.9,2.5)
\put(1.8,2.6){$K(\ell)d\ell$}
%\put(2.6,1.8){$\vec T_n(\ell)$}
\put(2.38,3.1){$- \vec T_n(\ell+d\ell)$}
\put(.8,1.9){\makebox[0cm][r]{$\tau(\ell)$}}
\put(.7,2.4){\makebox[0cm][r]{$-\tau(\ell+d\ell)$}}
\put(1.6,0){\large (a)}
\end{pspicture}
%%%%%%%%%%%%%%%%%% PS PICTURE
}%
\qquad
\qquad
\framebox{%
%%%%%%%%%%%%%%%%%% PS PICTURE: T_n balancing
\begin{pspicture}(-.07\textwidth,0)(.33\textwidth,.33\textwidth)
%%%%---- unit lengths for Latex, pstricks
\psset{unit=.1\textwidth}
\setlength{\unitlength}{.1\textwidth}
%\psgrid[subgriddiv=10,griddots=10,gridlabels=10pt]
%%%%---- Main curve
\pscurve[linewidth=.3mm,showpoints=false]{-}%
(0.6,0.6)(.93,1.5)(1,2)(.9,2.5)(.6,2.85)(0,3)
%%%%---- Lines
\psline[linewidth=.3mm]{->}%
(1,2)(1,3)
\psline[linewidth=.3mm]{->}%
(1,2)(2,2)
\psline[linestyle=dashed]{-}%
(.9,2.5)(2.5,2.5)
\psline{->}%
(1,2)(-.6,2)
\psline{->}%
(.9,2.5)(2.27,3.2)
\psline{->}%
(.9,2.5)(.5,3.3)
\psline{<-}%
(1.05,2.25)(1.4,2.35)
%%%%---- Positive direction for torques
\psarc{->}(1.4,1.2){.2}{-180}{90}
%%%%---- Actual directions for torques, angle
%\psarcn{->}(1,2){.1}{90}{-180}
%\psarc{->}(.9,2.5){.1}{0}{270}
\psarc{->}(.9,2.5){.5}{0}{26}
%%%%---- Labels, dots
\put(1.33,1.16){$+$}
\put(1.1,3){$\th$}
\put(1.95,1.8){$\nh$}
\psdots(1,2)(.9,2.5)
\put(1.8,2.6){$K(\ell)d\ell$}
\put(-.65,1.7){$\vec T_n(\ell)$}
\put(2.38,3.1){$- \vec T_n(\ell+d\ell)$}
\put(.4,3.2){\makebox[0cm][r]{$-\vec T_t(\ell + d\ell)$}}
\put(1.3,2.2){$\sigma$}
\put(1.2,0){\large (b)}
%\put(.8,1.9){\makebox[0cm][r]{$\tau(\ell)$}}
%\put(.7,2.4){\makebox[0cm][r]{$-\tau(\ell+d\ell)$}}
\end{pspicture}
%%%%%%%%%%%%%%%%%% PS PICTURE
}%
\qquad
\qquad
\framebox{%
%%%%%%%%%%%%%%%%%% PS PICTURE: T_t balancing
\begin{pspicture}(0,0)(.33\textwidth,.33\textwidth)
%%%%---- unit lengths for Latex, pstricks
\psset{unit=.1\textwidth}
\setlength{\unitlength}{.1\textwidth}
%\psgrid[subgriddiv=10,griddots=10,gridlabels=10pt]
%%%%---- Main curve
\pscurve[linewidth=.3mm,showpoints=false]{-}%
(0.6,0.6)(.93,1.5)(1,2)(.9,2.5)(.6,2.85)(0,3)
%%%%---- Lines
\psline[linewidth=.3mm]{->}%
(1,2)(1,3)
\psline[linewidth=.3mm]{->}%
(1,2)(2,2)
\psline[linestyle=dashed]{-}%
(.9,2.5)(2.5,2.5)
%\psline{->}%
%(1,2)(2.6,2)
\psline{->}%
(.9,2.5)(2.27,3.2)
\psline{->}%
(1,2)(1,3.5)
\psline{->}%
(.9,2.5)(1.4,1.4)
%%%%---- Positive direction for torques
\psarc{->}(1.5,1.1){.2}{-180}{90}
%%%%---- Actual directions for torques, angle
%\psarcn{->}(1,2){.1}{90}{-180}
%\psarc{->}(.9,2.5){.1}{0}{270}
\psarc{->}(.9,2.5){.5}{0}{26}
%%%%---- Labels, dots
\put(1.43,1.06){$+$}
\put(1.1,3){$\th$}
\put(1.95,1.8){$\nh$}
\psdots(1,2)(.9,2.5)
\put(1.8,2.6){$K(\ell)d\ell$}
\put(2.38,3.1){$- \vec T_n(\ell+d\ell)$}
\put(.9,3.25){\makebox[0cm][r]{$\vec T_t(\ell)$}}
\put(1.5,1.45){$-\vec T_t(\ell+ d\ell)$}
\put(1.6,0){\large (c)}
%\put(2.6,1.8){$\vec T_n(\ell)$}
%\put(.8,1.9){\makebox[0cm][r]{$\tau(\ell)$}}
%\put(.7,2.4){\makebox[0cm][r]{$-\tau(\ell+d\ell)$}}
\end{pspicture}
}%
}%
\capitem{Balancing of (a) torques, (b) normal forces and (c)
tangential forces on a segment of the loop. $K$ and $K'$ are
sketched positive, the resulting directions of torques and forces
are shown. For $T_t$ this data is not enough, we have assumed
additionally $K^2>\mu$ --- this puts the center of the loop 
somewhere towards the left of each picture.
}
%%%%%%%%%%%%%%%%%% PS PICTURE
%\end{center}
\label{torqfig}
\end{figure}
%%%%%%%%%%%%%%%%%%%%%%%%%%%%%%% FIGURE
The equilibrium of
normal forces on the segment gives
\ble{nforceeq}
T_n(\ell)   - T_n(\ell + d\ell)
+ T_t(\ell + d \ell) K(\ell)d\ell -\sigma d\ell =0
\qquad \Rightarrow \qquad
-T'_n +KT_t-\sigma=0
\, ,
\ee
while the tangential components give
\ble{tforceeq}
T_t(\ell) - T_t(\ell + d\ell) - T_n(\ell + d \ell) K(\ell) d\ell =0
\qquad \Rightarrow \qquad
T'_t + KT_n =0
\, .
\ee
Notice how, in~(\ref{nforceeq}) and~(\ref{tforceeq}), the
curvature $K$ is responsible for the tangential force at $\ell + d\ell$
contributing a normal component at $\ell$ and {\em vice-versa}.
Solving the above system of equations we find
\fle{summtf}{%
\tau = -2K
\, , \qquad \qquad
T_n=-2K'
\, , \qquad \qquad
T_t=K^2-\mu
\, ,
}%
where, in the third relation, the integration constant was fixed to
the value $\mu$ by the requirement that one recover the differential 
equation for $K$, Eq.~(\ref{eq:D1}). We see from~(\ref{summtf})
that $-\mu$ is the tension of the loop at its inflection points,
if any. Notice also how~(\ref{eq:D1}) is obtained upon substitution 
of the last two of~(\ref{summtf}) in~(\ref{nforceeq}), thereby
identifying the physical origin of each of the terms in the
former. In a forthcoming publication, where we extend our
considerations to the case of a loop in space, it is shown how
the above expressions for the forces and torque follow, in a
model independent way, from an application of Noether's theorem.

Our submerged cylinder model for the loop leaves no room for
self-intersecting configurations, could there be any use for
these? Imagine the loop made of superconducting material, in
the presence of a uniform magnetic field perpendicular to its
plane. In the limit where the magnetic flux due to
self-inductance is negligible compared to the one due to the
external field, the area of the loop has to be constant to keep
the flux constant. One can
then adjust the area by changing the magnetic field and, for small
enough areas, self-intersecting configurations will appear. 
%%%%%%%%%%%%%%%%%%%%%%%%%%%%%%%%%%%%%%%%%%%%%%%%%%%%%%%%%%%%%%%%%
%%%%%%%%%%%%%%%%%%%%%%%%%%%%%%%%%%%%%%%%%%%%%%%%%%%%%%%%%%%%%%%%%
\section{Configurations}
\label{Config}
%%%%%%%%%%%%%%%%%%%%%%%%%%%%%%%%%%%%%%%%%%%%%%%%%%%%%%%%%%%%%%%%%
%%%%%%%%%%%%%%%%%%%%%%%%%%%%%%%%%%%%%%%%%%%%%%%%%%%%%%%%%%%%%%%%%
\subsection{Qualitative remarks}
\label{qr}
%%%%%%%%%%%%%%%%%%%%%%%%%%%%%%%%%%%%%%%%%%%%%%%%%%%%%%%%%%%%%%%%%
For the purposes of this section, we may set, as explained above, 
$\sigma=1$ in the expression for
the potential $V(K)$. Its critical points are given by the zeros
of its derivative, \ie, \ by the roots $K_i$, $i=1,2,3$, of
\be
\label{Vpeqz}
K^3 - \mu K -1 = 0
\, .
\ee
We find
\be
\label{Vproots}
K_1 = -K_+ + i \sqrt{3} K_-
\, , \qquad
K_2 = -K_+ - i \sqrt{3} K_-
\, , \qquad
K_3 = 2K_+
\, ,
\ee
where 
\be
\label{mdef}
K_+ \equiv {m \over 12} + {\mu \over m}
\, , \qquad \qquad
K_- \equiv {m \over 12} - {\mu \over m}
\, , \qquad \qquad
m^3 \equiv 108 +108 \sqrt{1 - \frac{\mu^3}{\mu_0^3}}
\, ,
\ee
where $\mu_0 \equiv 3/2^{2/3}$.
Of these, $K_3$ is always real while the first two are real only
for $\mu \geq \mu_0$. 
Notice that, for any $\mu$,
the effect of the quadratic and the linear term in the
potential is relatively important only in a neighborhood of the 
origin. For
motions of the particle with sufficiently high energy $E$, the
time spent by the particle in this region is negligible and the
quartic term dominates the motion. As a result, we may conclude
that $\Theta_0$ approaches zero with increasing $E$, due to the
approximate symmetry $K \rightarrow -K$ of the motion. The term
linear in $K$ spoils this symmetry 
and, for $\sigma > 0$ (as taken in this section), makes 
$\Theta_0$ slightly positive for large $E$. 
As $\Theta_0$ descends, with increasing $E$, from 
this positive value to 0, it will cross all critical values
$2\pi/n$ for $n$ greater than some $n_0$. We expect therefore to 
encounter configurations with arbitrarily high $n$, for all
$\mu$ --- see  Fig.{}~\ref{Theta0Rpm}. 

\par
%%%%%%%%%%%%%%%%%% FIGURE
\setlength{\figurewidth}{.4\textwidth}
\begin{floatingfigure}{\figurewidth}
\rule{0mm}{.675\figurewidth}
\begin{pspicture}(0\figurewidth,0\figurewidth)%
                 (.9\figurewidth,.625\figurewidth)
\setlength{\unitlength}{.25\figurewidth}
\psset{xunit=.25\figurewidth,yunit=.25\figurewidth,arrowsize=1.5pt 3}
%\psgrid[subgriddiv=10,griddots=5,gridlabels=8pt]
%%%%%%% Puts
\put(3.1,1.42){\makebox[0cm][l]{$\mu/\mu_0$}}
\put(2,2.35){\makebox[0cm][r]{$\Theta_{0i}^{(max)}/2\pi$}}
\put(2.9,2.05){\makebox[0cm][l]{$i=3$}}
\put(2.9,.7){\makebox[0cm][l]{$i=1$}}
%%%%%%% Axes
\psline[linewidth=.3mm]{->}%
(2.068,.27)(2.068,2.5)
\psline[linewidth=.3mm]{->}%
(.35,1.65)(3.3,1.65)
\psline[linewidth=.15mm,linestyle=dashed]{-}%
(2.63,.27)(2.63,2.5)
\psline[linewidth=.15mm]{-}%
(2.068,2.32)(3.2,2.32)
\psline[linewidth=.15mm]{-}%
(2.068,.96)(3.2,.96)
%%%%%%% Included graphic
\raisebox{\totalheight}{%
\includegraphics[angle=270,width=.9\figurewidth]{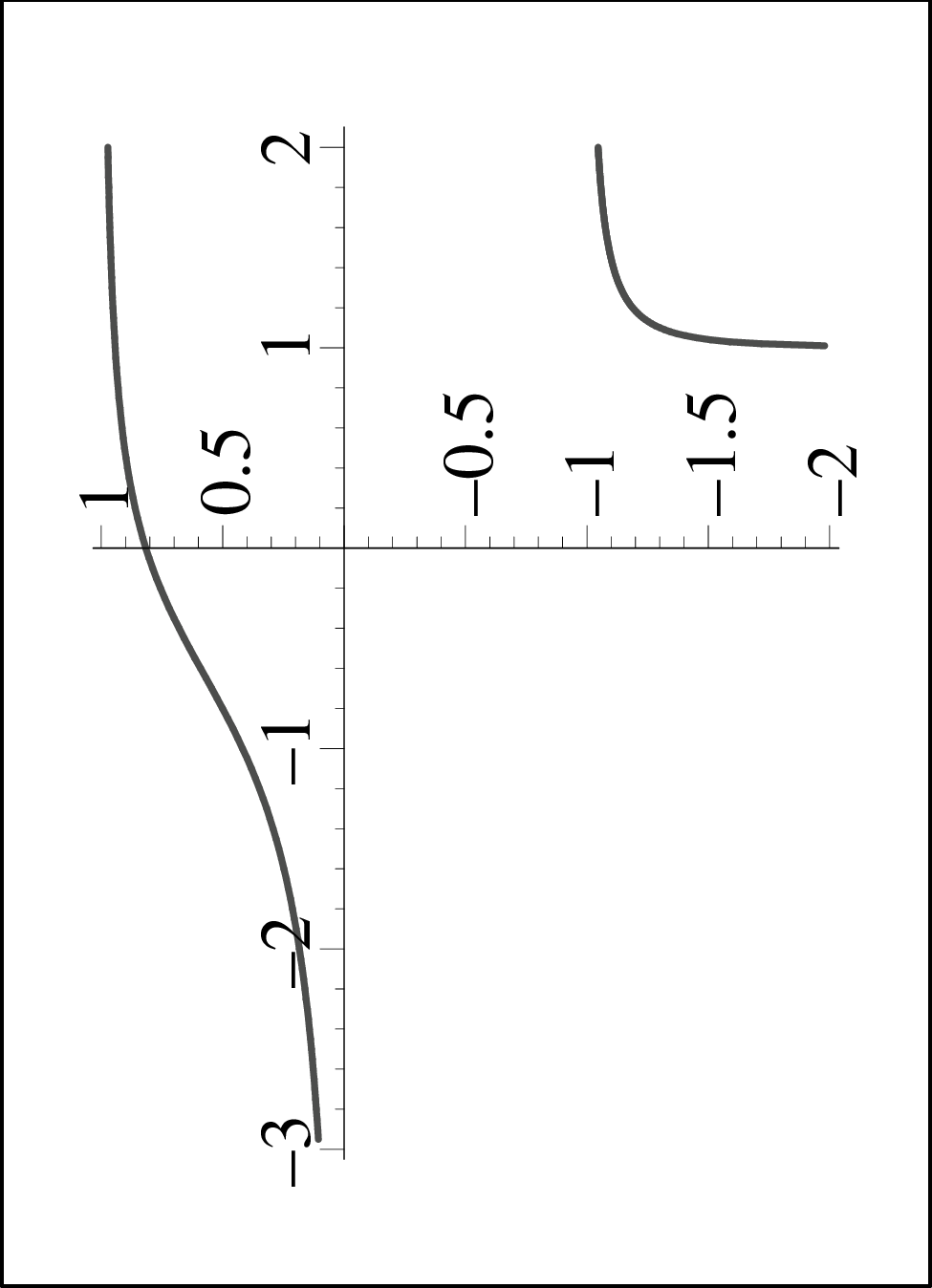}%
}
\end{pspicture}%
\capitem{$\Theta_{0i}^{max}(\mu)$, $i=1,3$ \\ \\ }
\label{the0max}
\end{floatingfigure}
%%%%%%%%%%%%%%%%%% FIGURE
\par

%%%%%%%%%%%%%%%%%% FIGURE
\setlength{\figurewidth}{.8\textwidth}
\begin{figure*}
%\rule{0mm}{.675\figurewidth}
\begin{pspicture}(0\figurewidth,0\figurewidth)%
                 (\textwidth,.7\figurewidth)
\setlength{\unitlength}{.25\figurewidth}
\psset{xunit=.25\figurewidth,yunit=.25\figurewidth,arrowsize=1.5pt 3}
%\psgrid[subgriddiv=10,griddots=5,gridlabels=8pt]
%%%%%%% Puts
\put(3.38,2.1){\makebox[0cm][l]{$\bar{\mu} \! = \! 2$}}
\put(2.68,2.1){\makebox[0cm][l]{$\bar{\mu} \! = \! 1.5$}}
\put(2.13,2.1){\makebox[0cm][l]{$\bar{\mu} \! = \! 1.01$}}
\put(1.74,2){\makebox[0cm][l]{$\bar{\mu} \! = \! 0.5$}}
\put(1.2,1.7){\makebox[0cm][l]{$\bar{\mu} \! = \! -1$}}
\put(1.15,1.47){\makebox[0cm][l]{$\bar{\mu} \! = \! -2.5$}}
\put(4.1,1.25){\makebox[0cm][l]{$\bar{E}$}}
\put(1.1,2.43){\makebox[0cm][l]{$\Theta_{03}/2\pi$}}
\put(3.97,1.8){\makebox[0cm][l]{$n \! = \! 2$}}
\put(3.97,1.63){\makebox[0cm][l]{$n \! = \! 3$}}
\put(3.97,1.53){\makebox[0cm][l]{$n \! = \! 4$}}
\put(3.97,1.025){\makebox[0cm][l]{$n \! = \! -4$}}
\put(3.97,.92){\makebox[0cm][l]{$n \! = \! -3$}}
\put(3.97,.76){\makebox[0cm][l]{$n \! = \! -2$}}
\put(1.95,1.08){\makebox[0cm][r]{%
                $\bar{E}^{(2)}(\bar{\mu} \! = \! 0.5)$}}
\rput(1.97,1.1){\rnode{A}{}}
\rput(1.8,1.28){\rnode{B}{}}
\nccurve[angleA=45,angleB=-40,linewidth=.18mm]{->}{A}{B}
\psdots[dotscale=1]%
(1.79,1.825)
\put(1.82,1.828){\makebox[0cm][l]{$A$}}
\psdots[dotscale=1]%
(1.79,1.3)
\psdots[dotscale=1]%
(2.14,1.04)
\put(2.13,1){\makebox[0cm][r]{$B_1$}}
\psdots[dotscale=1]%
(2.195,1.04)
\put(2.22,1){\makebox[0cm][l]{$B_2$}}
%%%%%%% Axes
\psline[linewidth=.3mm]{->}%
(1.065,1.3)(4.05,1.3)
\psline[linewidth=.3mm]{->}%
(1.065,0.26)(1.065,2.45)
\psline[linewidth=.15mm,linestyle=dashed]{-}%
(1.79,1.825)(1.79,1.3)
%%%%%%% Included graphic
\centerline{%
\raisebox{\totalheight}{%
\includegraphics[angle=270,width=.9\figurewidth]{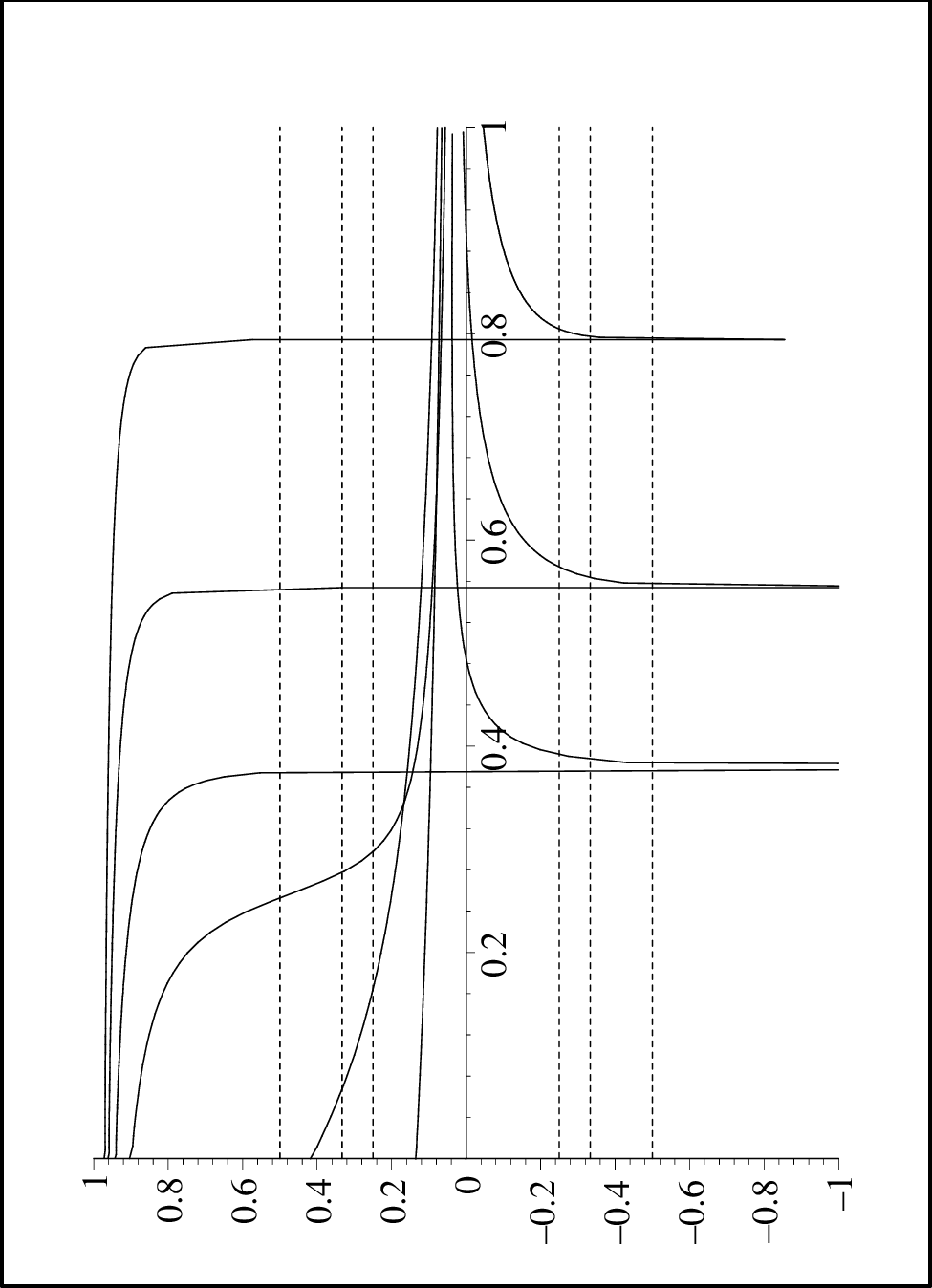}%
}%
}
\end{pspicture}%
%%%%%%% Caption
\capitem{$\Theta_{03}(\bar{E})$, $\bar{E} \equiv
(E-V(K_3))/\mu_0^2$, for various values of $\bar{\mu} \equiv
\mu/\mu_0$. Also shown, with horizontal dashed lines, are some of 
the values of
$\Theta_{03}$ that give rise to closed configurations. The
corresponding values of $\bar{E}$ can be read off as in the case
$\bar{\mu} \! = \! 0.5$, $n \! = \! 2$ shown --- the
(self-intersecting) configuration itself appears in
Fig.{}~\ref{configint}. Notice how the
$\bar{\mu}=-1$, $-2.5$ curves miss the $n=2$, $n=2,3,4$
configurations respectively.}
\label{Theta0Rpm}
\end{figure*}
%%%%%%%%%%%%%%%%%% FIGURE

On the other extreme, when the energy $E$ is only slightly greater
than a local minimum of the potential (at $K_1$ or $K_3$),  and 
the particle
oscillates around that minimum, we may approximate $V(K)$
by a quadratic expression in $K$ and find for $\Theta_0$ the
limiting value
\be
\label{the0min}
\Theta_{0i}^{(max)} = 2\pi
\frac{\sqrt{2}}{\sqrt{3-\frac{\mu}{{K_i}^2}}}
\, , \qquad \qquad
i=1,3
\, ,
\ee
where the superscript {\em (max)} is used because, as we shall
see shortly, this is actually a maximum of
$\Theta_{0i}(\mu,E)$, for fixed $\mu$. Starting
from~(\ref{Vproots}), one infers the limiting values
\ble{The1lim}
\Theta_{01}^{(max)}(\mu \to \mu_0^+)=-\infty
\, , \qquad 
\Theta_{01}^{(max)}(\mu \to \infty)=-2\pi
\, , 
\ee
as well as
\ble{The3lim}
\Theta_{03}^{(max)}(\mu \to -\infty)=0
\, , \qquad 
\Theta_{03}^{(max)}(\mu = 0)=2\pi \sqrt{{2 \over 3}}
\, , \qquad 
\Theta_{03}^{(max)}(\mu \to \infty)=2\pi
\, .  
\ee
A plot of 
$\Theta_{0i}^{(max)}(\mu)$ is given in Fig.{}~\ref{the0max}. We
notice that, for $\mu > \mu_0$, {\em there are no non-self-intersecting 
configurations corresponding to oscillations in the left well}.   

We explain now why the value $n=1$ was omitted
in~(\ref{condclos}). The reasoning behind that relation was that
a closed configuration should correspond to $n$ complete
oscillations of the particle. This is not necessarily so in the
case of the circle. {\em Any} number of ``oscillations'' (of zero
amplitude) will fit into a circle, including irrational numbers,
and indeed,~(\ref{the0min}) shows that the circle corresponding
to the particle resting at $K_i$, $i=1,3$, is made of
$\sqrt{3-\frac{\mu}{{K_i}^2}}/\sqrt{2}$ complete oscillations.

To examine what happens for values of $E$ that render the
quadratic and linear terms important, we analyze the cases $\mu <
\mu_0$ and $\mu \geq \mu_0$ separately.
%%%%%%%%%%%%%%%%%%%%%%%%%%%%%%%%%%%%%%%%%%%%%%%%%%%%%%%%%%%%%%%%%
\subsection{\mathversion{bold} $\mu < \mu_0$ \mathversion{normal}}
\label{mulz}
%%%%%%%%%%%%%%%%%%%%%%%%%%%%%%%%%%%%%%%%%%%%%%%%%%%%%%%%%%%%%%%%%
The potential possesses only 
one minimum,
at $K_3 > 0$ given by~(\ref{Vproots}) (see Fig.~\ref{phase}). 
$\Theta_{03}$ starts at
$\Theta_{03}^{(max)}$, for $E=V(K_3)$, and for $\mu$ far from 
$\mu_0$, decreases monotonically 
to zero with increasing $E$. The configurations that do not 
appear are those for
which $2 \leq n < 2\pi/\Theta_0^{(max)}$ as well as all with
negative $n$. This condition determines a
set $\{\mu_n | n=2,3, \dots\}$ of critical values of $\mu$, such
that, for $\mu < \mu_n$, all configurations with $n'=2,3,\dots,n$
are absent (the circle is, of course, always there). 
As $\mu$ approaches $\mu_0$, the left wall of the potential
develops a plateau that tends to the horizontal, as $\mu \to
\mu_0$. The particle spends a relatively long time in this
(negative $K$) region, which results in a negative bump in the
$\Theta_0$ curve. As $\mu \to \mu_0$, the minimum of this bump
tends to $-\infty$ --- negative $n$ configurations appear
accordingly. Direct evaluation of the integral~(\ref{Thetaint})
gives~\cite{Byr.Fri:54}
\ble{Thetaexp}
\Theta_{03}(\mu,E)=\frac{8g(aB+bA)}{A-B} 
\left[ \alpha_2 \K (k) + \frac{\alpha - \alpha_2}{1-\alpha^2}
\Pi(\frac{\alpha^2}{\alpha^2-1},k) \right]
\, ,
\ee
where $k^2=\frac{(a-b)^2-(A-B)^2}{4AB}$,
\ble{vardef1}
\begin{array}{rclcrclcrcl}
\dss g \fedss \frac{1}{\sqrt{AB}}
\aqqadss
\alpha \fedss \frac{A-B}{A+B}
\aqqadss
\alpha_2 \fedss \frac{bA-aB}{aB+bA}
\\
\dss A^2 \fedss (a-b_1)^2+a_1^2
\aqqadss
B^2 \fedss (b-b_1)^2+a_1^2
\aqqadss
b_1 \fedss -\frac{1}{2}(a+b)
\\
\dss a_1^2 \fedss \frac{4}{a+b} + ab - \frac{1}{4}(a+b)^2
\aqqadss
a \fedss K_{max}(\mu,E)
\aqqadss
b \fe K_{min}(\mu,E)
\end{array}
\ee
and $\K$, $\Pi$ are the complete elliptic integrals of the first
and third kind respectively, given by
\ble{KPIdef}
\K(k) = \int_0^{\pi/2} \frac{d\theta}{\sqrt{1-k^2
\sin^2\theta}}
\, , \qquad \quad
\Pi(\alpha^2, k) = \int_0^{\pi/2} \frac{d \theta}{(1-\alpha^2
\sin^2 \theta) \sqrt{1-k^2 \sin^2 \theta}}
\, .
\ee
The result~(\ref{Thetaexp}) holds for $\mu > \mu_0$ as well, with
the appropriate choice of $K_{min}$, $K_{max}$. A plot of
$\Theta_{03}(\mu,E)$, for various values of $\mu$, appears in
Fig.{}~\ref{Theta0Rpm}.

A couple of remarks are in order at this point. 
Consider the point $A$ in Fig.{}~\ref{Theta0Rpm}, which is 
the intersection of
the $\bar{\mu}=0.5$ curve with the $n=2$ line, corresponding to
an $n=2$ configuration. Imagine now that $\bar{\mu}$ is
diminished continuously --- the corresponding curve will move
more or less downwards forcing $A$ to move to the left. The
corresponding $\bar{E}$ then diminishes which means that the
fictitious particle oscillates in the potential well with smaller 
amplitude. When $A$ hits the $\Theta_{03}$-axis, $\bar{E}$ is
zero, the particle sits at the bottom of the well and the
corresponding configuration becomes a circle. In other words,
{\em all
configurations like the one corresponding to the point $A$ can be
continuously deformed to a circle}. 

\par
%%%%%%%%%%%%%%%%%% FIGURE
\setlength{\figurewidth}{.45\textwidth}
\begin{floatingfigure}{\figurewidth}
\rule{0mm}{.685\figurewidth}
\begin{pspicture}(0\figurewidth,0\figurewidth)%
                 (.9\figurewidth,.625\figurewidth)
\setlength{\unitlength}{.25\figurewidth}
\psset{xunit=.25\figurewidth,yunit=.25\figurewidth,arrowsize=1.5pt 3}
%\psgrid[subgriddiv=10,griddots=5,gridlabels=8pt]
%%%%%%% Puts
%\put(3.1,1.42){\makebox[0cm][l]{$\mu/\mu_0$}}
%\put(2,2.35){\makebox[0cm][r]{$\Theta_{0i}^{(max)}/2\pi$}}
%\put(2.9,2.05){\makebox[0cm][l]{$i=3$}}
%\put(2.9,.7){\makebox[0cm][l]{$i=1$}}
%%%%%%% Axes
\psline[linewidth=.3mm]{->}%
(0,1)(3.3,1)
\psline[linewidth=.3mm]{->}%
(0,.1)(0,2.5)
\psline[linewidth=.15mm,linestyle=dashed]{-}%
(0,2)(3.3,2)
\psline[linewidth=.15mm,linestyle=dashed]{-}%
(0,1.2)(3.3,1.2)
\psline[linewidth=.15mm,linestyle=dashed]{-}%
(0,.75)(3.3,.75)
\pscurve[linewidth=.25mm]{-}%
(0,1.9)(.5,1.89)(1,1.8)(1.05,1.6)(1.1,.75)%
(1.15,.65)(1.2,.75)(1.33,1.2)%
(2,1.4)(3.2,1.1)
\psdots[dotscale=1]%
(1.1,.75)(1.2,.75)(1.07,1.2)(1.32,1.2)(2.92,1.2)
\put(1.02,1.1){\makebox[0cm][r]{$B'_1$}}
\put(1.38,1.1){\makebox[0cm][l]{$B'_2$}}
\put(2.92,1.3){\makebox[0cm][l]{$B'_3$}}
\put(1.07,.56){\makebox[0cm][r]{$B_1$}}
\put(1.25,.56){\makebox[0cm][l]{$B_2$}}
\put(.1,2.05){\makebox[0cm][l]{$n \! = \! 1$}}
\put(.1,1.25){\makebox[0cm][l]{$n \! = \! 5$}}
\put(.1,.8){\makebox[0cm][l]{$n \! = \! -4$}}
\put(.05,2.5){\makebox[0cm][l]{$\Theta_{03}/2\pi$}}
\put(3.35,.95){\makebox[0cm][l]{$\bar{E}$}}
\end{pspicture}%
\capitem{Bifurcations: as $\bar{\mu}$ is reduced, the minimum of
the curve shifts upwards and $B_1$, $B_2$ collapse to a 
point. Further decrease of $\bar{\mu}$ forces $B'_1$, $B'_2$ to
collapse to a point. The corresponding configurations cannot be
continuously deformed to a circle.% 
}
\label{exag}
\end{floatingfigure}
%%%%%%%%%%%%%%%%%% FIGURE
\par

This is not the case though with
the points $B_1$, $B_2$ in the same figure. Consider $B_1$ --- 
it corresponds
to an $n=-4$ configuration for $\bar{\mu}=1.01$. Imagine now that
$\bar{\mu}$ is diminished in a continuous way. As soon as it
becomes smaller then 1, the infinite negative pole of the
corresponding curve is softened to a negative minimum which, for
$\bar{\mu}$ sufficiently close to 1, still intersects the $n=-4$
line. Given that the $\mu$-curves in Fig.~\ref{Theta0Rpm} become
almost horizontal at high $E$, we give an exagerrated sketch of
the situation in Fig.~\ref{exag}. 
As the value of $\bar{\mu}$ is lowered, this negative
minimum rises and, for a critical value of $\bar{\mu}$, will
just touch the $n=-4$ line, \ie, \ $B_1$ and $B_2$
collapse to a single point $B$. The configuration that corresponds to
this point is not a circle, since $\bar{E}$ is positive and the
particle oscillates with a finite amplitude, \ie, \ $K$ is not
constant. We conclude that {\em configurations corresponding to points
like $B_1$, $B_2$ cannot be continuously deformed to a circle}.
Moreover, {\em there exist bifurcation points, like $B$ above, 
distinct from the circle.} 
When $\bar{\mu}$ is lowered even more, the minimum of the
$\bar{\mu}$-curve becomes positive and remarks similar to the
above can be made about its points of intersection $B'_1$, $B'_2$
with sufficiently high $n$ lines (see Fig.~\ref{exag}). 
%%%%%%%%%%%%%%%%%%%%%%%%%%%%%%%%%%%%%%%%%%%%%%%%%%%%%%%%%%%%%%%%%
\subsection{\mathversion{bold} $\mu \geq \mu_0$ \mathversion{normal}}
\label{mugez}
\begin{figure}
%%%%%%%%%%%%%%%% PSPICTURE
\setlength{\figurewidth}{.43\textwidth}
%\fbox{%
\begin{pspicture}(0\figurewidth,0\figurewidth)%
                 (.48\figurewidth,.625\figurewidth)
\setlength{\unitlength}{.25\figurewidth}
\psset{xunit=.25\figurewidth,yunit=.25\figurewidth,arrowsize=1.5pt 3}
%\psgrid[subgriddiv=10,griddots=5,gridlabels=8pt]
%%%%%%% Puts
\put(3.3,1.5){\makebox[0cm][l]{$\bar{K}$}}
\put(1.8,2.35){\makebox[0cm][l]{$\bar{V}$}}
%%%%%%% Axes
\psline[linewidth=.2mm]{->}%
(.35,1.61)(3.25,1.61)
\psline[linewidth=.2mm]{->}%
(1.7,.25)(1.7,2.4)
%%%%%%% Included graphic
\raisebox{\totalheight}{%
\includegraphics[angle=270,width=.9\figurewidth]{potV4.ps}%
}
\end{pspicture}%
%}
%%%%%%%%%%%%%%%% PSPICTURE
\hfill
%%%%%%%%%%%%%%%% PSPICTURE
\setlength{\figurewidth}{.43\textwidth}
%\fbox{%
\begin{pspicture}(0\figurewidth,.05\figurewidth)%
                 (.9\figurewidth,.625\figurewidth)
\setlength{\unitlength}{.25\figurewidth}
\psset{xunit=.25\figurewidth,yunit=.25\figurewidth,arrowsize=1.5pt 3}
%\psgrid[subgriddiv=10,griddots=5,gridlabels=8pt]
%%%%%%% Puts
\put(2.7,.3){\makebox[0cm][l]{$X$}}
\put(1.65,2.5){\makebox[0cm][r]{$Y$}}
%%%%%%% Axes
\psline[linewidth=.2mm]{->}%
(1,.48)(2.72,.48)
\psline[linewidth=.2mm]{->}%
(1.75,.45)(1.75,2.5)
%%%%%%% Included graphic
\raisebox{\totalheight}{%
\includegraphics[angle=270,width=.96\figurewidth]{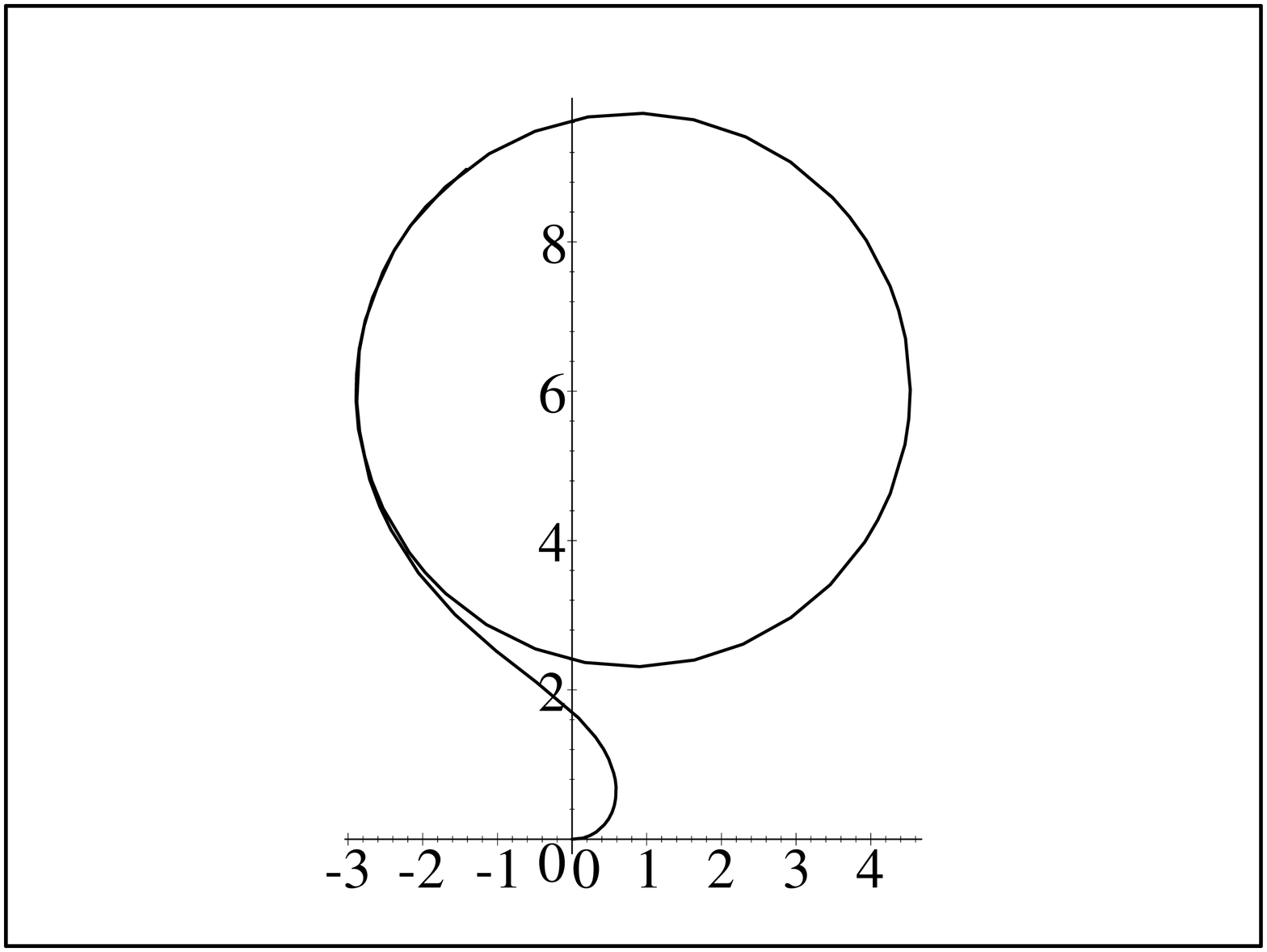}%
}
\end{pspicture}%
%}
%%%%%%%%%%%%%%%% PSPICTURE
%%%%%%%%%%%%%%%% CAPTION
\capitem{%
On the left:
the potential $\bar{V}(\bar{K})$ for $\bar{\mu}=2$.
On the right:
spiralling configuration for $E=V(K_2)$, 
with the particle starting at the turning point on the right with
zero velocity.
}
\label{potspi}
\end{figure}

%%%%%%%%%%%%%%%%%%%%%%%%%%%%%%%%%%%%%%%%%%%%%%%%%%%%%%%%%%%%%%%%%
\noindent The potential posseses two
local minima, at $K_1 < 0$ and $K_3 > 0$ (with $V(K_1) > 
V(K_3)$) and one local maximum at
$K_2 < 0$ (see Fig.{}~\ref{potspi}). 
Depending on its energy and where it is started from, the
particle is confined in the left well, the right well or visits
both during every oscillation.

We note the following, regarding the asymptotic behaviour of
$\Theta_{0i}$ as $\mu \rightarrow \infty$. In this regime, the
linear term becomes negligible in $V(K)$ and the integral giving
$\Theta_{0i}$ can be easily seen to reduce (with $K^2 \rightarrow
x$) to an integral giving (half) the {\em period} of a {\em harmonic}
oscillator in a potential $x^2/4 - \mu x/2$, which is 
independent of its  energy $E$, 
as well as the `constant force' $\mu$. Another way to see the
$\mu$-independence is by observing that, with $\sigma =0$ (which is
equivalent to sending $\mu$ to infinity), one lies on the
$\mu$-axis in the $(\sigma,\mu)$-plane, which is an orbit under
scaling, hence changes in $\mu$ in this regime leave $\Theta_{0i}$
invariant. Another novelty in this case is that, as the
energy approaches $V(K_2)$ (either from above or below), the period 
of the motion
tends to infinity, most of which the particle spends at $K_2$.
Assume we start the particle at the turning point on the right
with zero velocity and with just the right energy to reach the
central maximum of $V$, $E=V(K_2)$. 
The corresponding curve starts in a counterclockwise sense with
decreasing curvature, passes through an inflection point and
acquires infinite length  spiralling forever
clockwise as it approaches asymptotically a circle of radius
$1/K_2$, see Fig.{}~\ref{potspi}.

%%%%%%%%%%%%%%%%%%%%%%%% FIGURE
%%%%%%%%%%%%%%%%%%%%%%%% FIGURE
$\Theta_{03}$ tends accordingly to $-\infty$, this corresponds to
the negative poles in Fig.{}~\ref{Theta0Rpm}. As mentioned
already, for $V(K_1) <E< V(K_2)$, 
$\Theta_0$ will have two branches, $\Theta_{01}$ and
$\Theta_{03}$, corresponding to
the motions confined to either well. Both of these branches will
tend to $-\infty$ as $E \rightarrow V(K_2)$. For $E> V(K_2)$ 
there is only one branch, which starts from $-\infty$
and approaches asymptotically zero (as $E \rightarrow \infty$),
after having reached a positive local maximum. 
A three-dimensional plot of $\Theta_{03}(\mu,E)$
is given in Fig.{}~\ref{T0Rpm}.
We present a representative collection of configurations in 
Fig.{}~\ref{config}. We also show, for reference purposes, some 
self-intersecting configurations in 
Fig.{}~\ref{configint}. 

%%%%%%%%%%%%%%%%%% FIGURE
\setlength{\figurewidth}{.8\textwidth}
\begin{figure*}
%\rule{0mm}{.675\figurewidth}
\begin{pspicture}(0\figurewidth,0\figurewidth)%
                 (\textwidth,.7\figurewidth)
\setlength{\unitlength}{.25\figurewidth}
\psset{xunit=.25\figurewidth,yunit=.25\figurewidth,arrowsize=1.5pt 3}
%\psgrid[subgriddiv=10,griddots=5,gridlabels=8pt]
%%%%%%% Puts
\put(.83,1.27){\makebox[0cm][r]{$\bar{\mu}$}}
\put(4.1,1.27){\makebox[0cm][l]{$\bar{E}$}}
\put(2.41,2.67){\makebox[0cm][r]{$\Theta_{03}/2\pi$}}
%%%%%%% Axes
%\psline[linewidth=.3mm]{->}%
%(2.068,.27)(2.068,2.5)
%\psline[linewidth=.15mm,linestyle=dashed]{-}%
%(2.63,.27)(2.63,2.5)
%%%%%%% Included graphic
\centerline{%
\raisebox{\totalheight}{%
\includegraphics[angle=270,width=\figurewidth]{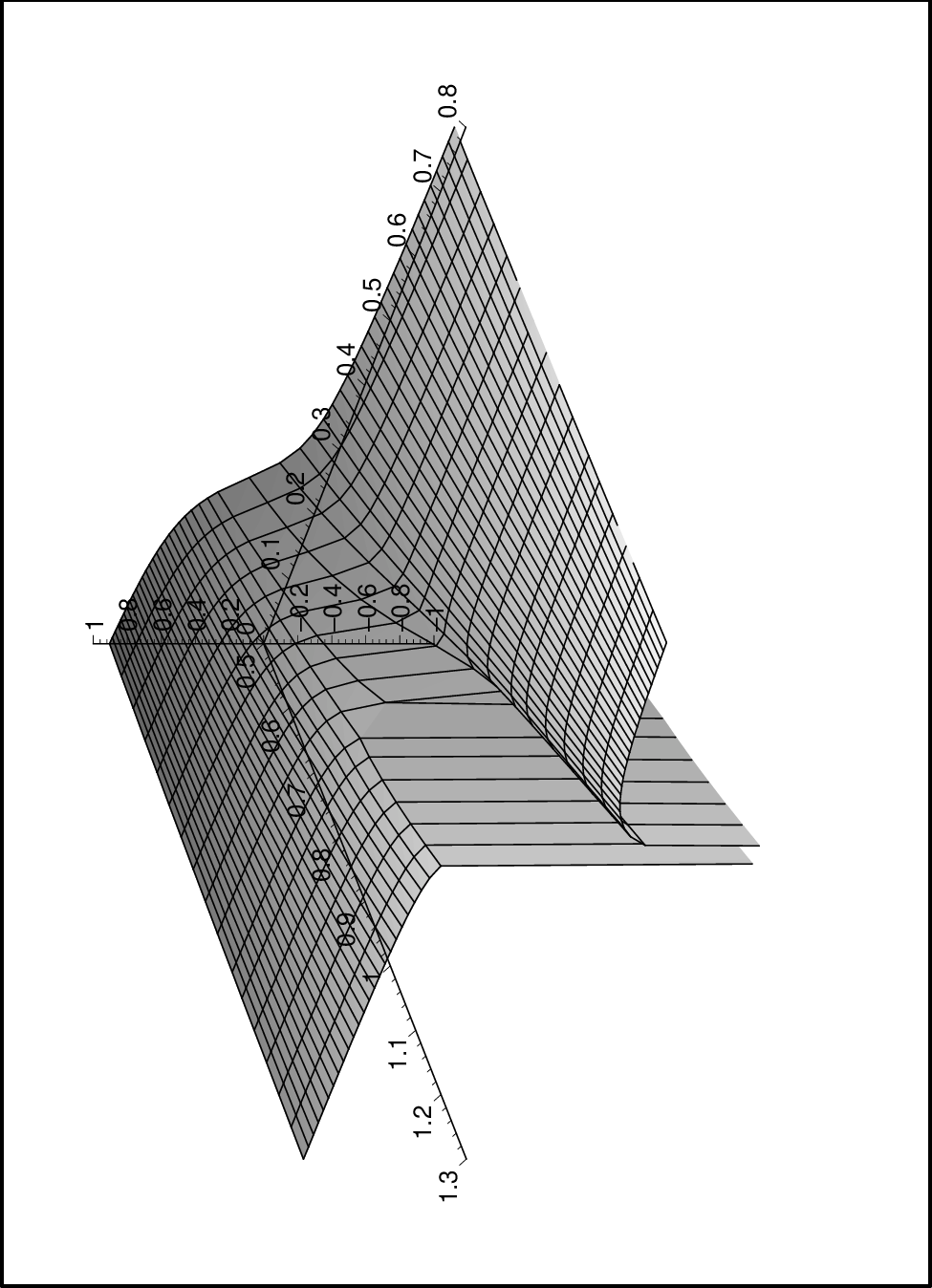}%
}%
}
\end{pspicture}%
\capitem{The angle $\Theta_{03}(\bar{\mu}, \bar{E})$, for $0.5 \leq
\bar{\mu} \leq 1.3$, in steps of 0.05 and $0 \leq \bar{E} \leq
.8$, in variable step. The $\bar{\mu}=1$ curve
has been shifted to $\bar{\mu}=1.01$ in order to avoid the
ambiguity in the number of roots of $\dot{V}$. The negative pole
in the curves for $\bar{\mu} > 1$ occurs at $E=V(K_2)$.%
}
\label{T0Rpm}
\end{figure*}
%%%%%%%%%%%%%%%%%% FIGURE

%%%%%%%%%%%%%%%%%% FIGURE
\setlength{\figurewidth}{.8\textwidth}
\begin{figure*}
%\rule{0mm}{.675\figurewidth}
\begin{pspicture}(.1\figurewidth,0\figurewidth)%
                 (\textwidth,.85\figurewidth)
\setlength{\unitlength}{.25\figurewidth}
\psset{xunit=.25\figurewidth,yunit=.25\figurewidth,arrowsize=1.5pt 3}
%\psgrid[subgriddiv=10,griddots=5,gridlabels=8pt]
%%%%%%% Puts
%\put(.83,1.27){\makebox[0cm][r]{$\bar{\mu}$}}
%\put(4.1,1.27){\makebox[0cm][l]{$\bar{E}$}}
%\put(2.43,2.65){\makebox[0cm][r]{$\Theta_{03}$}}
%%%%%%% Axes
%\psline[linewidth=.3mm]{->}%
%(2.068,.27)(2.068,2.5)
%\psline[linewidth=.15mm,linestyle=dashed]{-}%
%(2.63,.27)(2.63,2.5)
%%%%%%% Included graphic
\centerline{%
\raisebox{-.53\totalheight}{%
$\qquad \qquad$ 
$\qquad$ 
\includegraphics[angle=0,width=1.9\figurewidth]{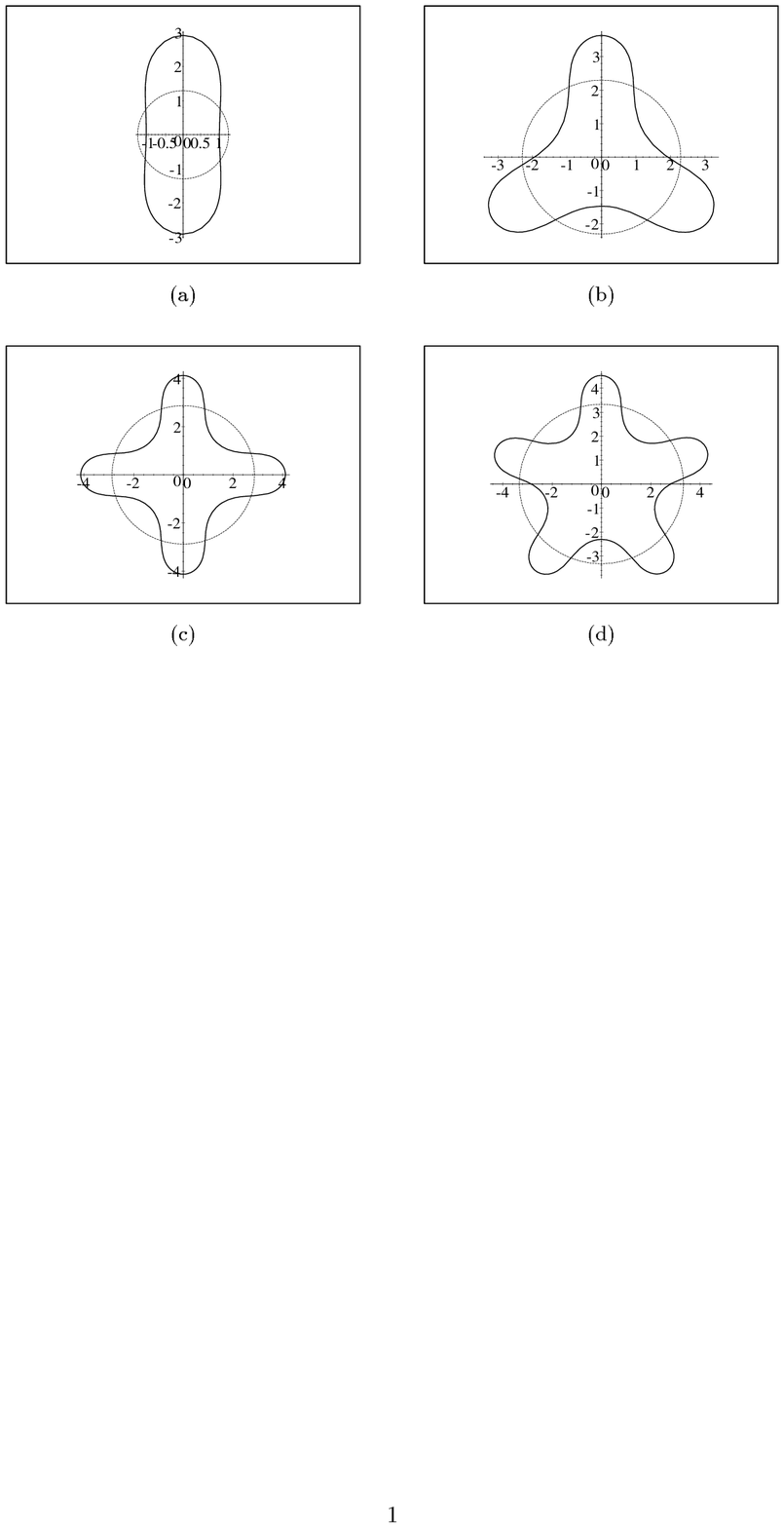}%
}%
}
\end{pspicture}%
\capitem{%
Non-self-intersecting configurations for $\bar{\mu}=-0.5$,
$\sigma=1$ and $n=2,3,4,5$.  Also shown, for each configuration, 
is the circle of radius
$X_0$ (see Eq.~(\ref{geomproof})), passing through its inflection 
points.
}
\label{config}
\end{figure*}
%%%%%%%%%%%%%%%%%% FIGURE

%%%%%%%%%%%%%%%%%% FIGURE
\setlength{\figurewidth}{.8\textwidth}
\begin{figure*}
%\rule{0mm}{.675\figurewidth}
\begin{pspicture}(.1\figurewidth,0\figurewidth)%
                 (\textwidth,.85\figurewidth)
\setlength{\unitlength}{.25\figurewidth}
\psset{xunit=.25\figurewidth,yunit=.25\figurewidth,arrowsize=1.5pt 3}
%\psgrid[subgriddiv=10,griddots=5,gridlabels=8pt]
%%%%%%% Puts
%\put(.83,1.27){\makebox[0cm][r]{$\bar{\mu}$}}
%\put(4.1,1.27){\makebox[0cm][l]{$\bar{E}$}}
%\put(2.43,2.65){\makebox[0cm][r]{$\Theta_{03}$}}
%%%%%%% Axes
%\psline[linewidth=.3mm]{->}%
%(2.068,.27)(2.068,2.5)
%\psline[linewidth=.15mm,linestyle=dashed]{-}%
%(2.63,.27)(2.63,2.5)
%%%%%%% Included graphic
\centerline{%
\raisebox{-.53\totalheight}{%
$\qquad \qquad$ 
$\qquad$ 
\includegraphics[angle=0,width=1.9\figurewidth]{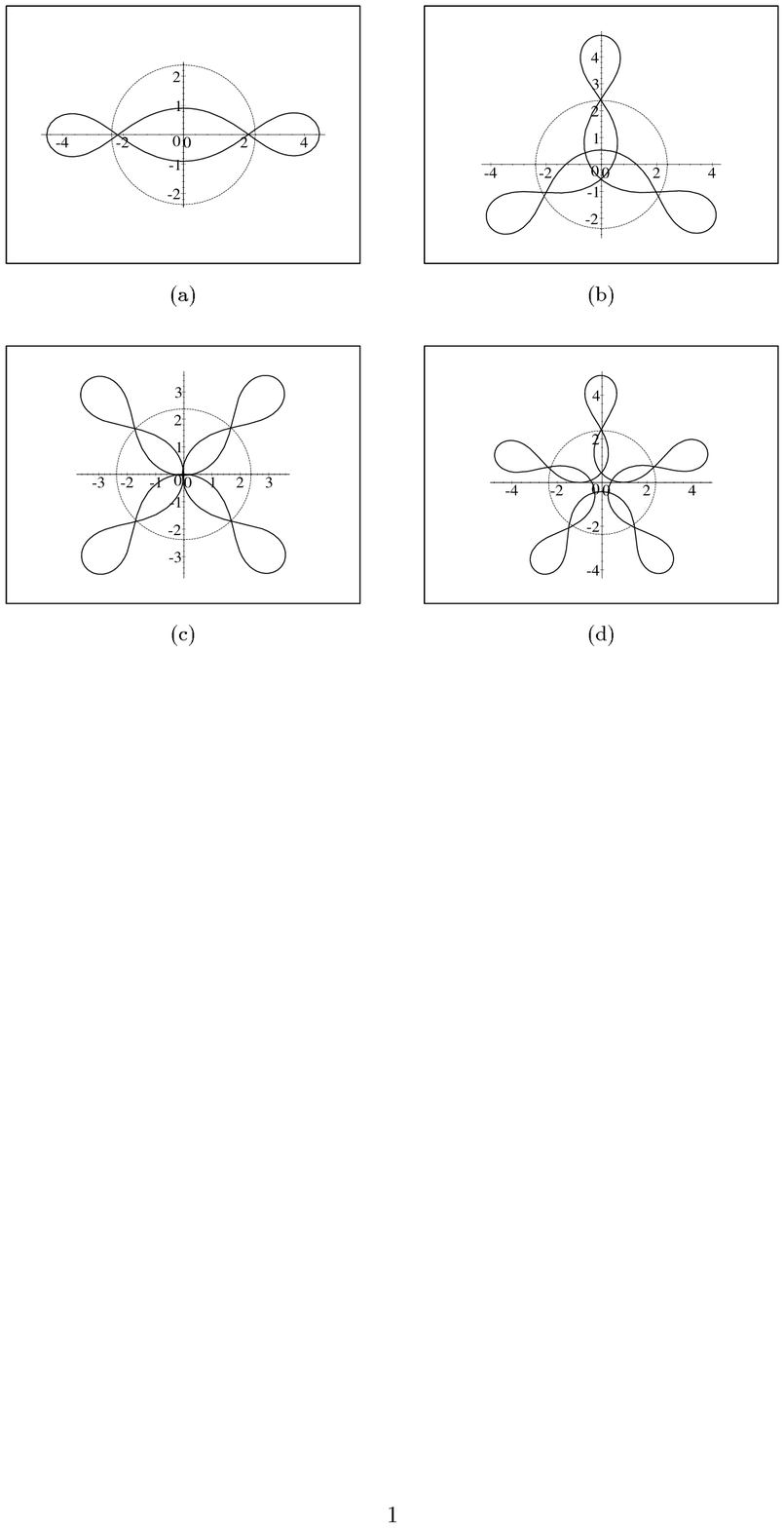}%
}%
}
\end{pspicture}%
\capitem{%
\textsf{Self-intersecting configurations for $\bar{\mu}=0.5$,
$\sigma=1$ and  $n=2,3,4,5$.  Also shown, for each configuration, 
is the circle of radius
$X_0$ (see Eq.~(\ref{geomproof})), passing through its inflection 
points.
}
}
\label{configint}
\end{figure*}
%%%%%%%%%%%%%%%%%% FIGURE

%%%%%%%%%%%%%%%%%%%%%%%%%%%%%%%%%%%%%%%%%%%%%%%%%%%%%%%%%%%%%%%%%
%%%%%%%%%%%%%%%%%%%%%%%%%%%%%%%%%%%%%%%%%%%%%%%%%%%%%%%%%%%%%%%%%
\section{Perturbations}
\label{PA}
%%%%%%%%%%%%%%%%%%%%%%%%%%%%%%%%%%%%%%%%%%%%%%%%%%%%%%%%%%%%%%%%
%%%%%%%%%%%%%%%%%%%%%%%%%%%%%%%%%%%%%%%%%%%%%%%%%%%%%%%%%%%%%%%%%
%%%%%%%%%%%%%%%%%%%%%%%%%%%%%%%%%%%%%%%%%%%%%%%%%%%%%%%%%%%%%%%%
\subsection{Equilibrium-connecting deformations}
%%%%%%%%%%%%%%%%%%%%%%%%%%%%%%%%%%%%%%%%%%%%%%%%%%%%%%%%%%%%%%%%
We study here the following problem: given an  
equilibrium configuration $\vec X$,
find an infinitesimal deformation $\deps \vec
X=\epsilon \nh$ such that $\vec X + \deps \vec X$
describes a nearby equilibrium. Notice that $\deps \vec X$ is not
necessarily the displacement vector of some point of the loop ---
the latter will generally have a tangential component as well.
 Our motivation is to
obtain new solutions from known ones, with the closure
condition automatically satisfied. 

Using some
of the formulae listed in the appendix, the linearization of the
equilibrium condition~(\ref{eq:D1}) gives 
that the deformation $\epsilon$ must satisfy 
\bae
\label{epsde4}
DE_4(\epsilon) 
& \equiv & 2 
\epsilon'''' +(5K^2-\mu)\epsilon'' +10KK' \epsilon'
+(12E-{5 \over 2} K^4 +6\mu K^2 +10\sigma K)\epsilon
\ff
 & = & 
-\deps \mu K - \deps \sigma
\, ,
\eae
where the deformed configuration satisfies~(\ref{eq:D1}),
with $\mu \to \mu + \deps \mu$, $\sigma \to \sigma +
\deps \sigma$. The energy  will change as well, $E \to E + \deps
E$. This appears explicitly in the
linearization of~(\ref{DEKalt}), which gives
\bae
\label{epsde3}
DE_3(\epsilon) 
& \equiv &
2 K' \epsilon''' + (K^3 - \mu K - \sigma ) \epsilon''
+ 2 K^2 K' \epsilon' + K ( 12 E - {1 \over 2} K^4
+ 2 \mu K^2 + 5 \sigma K ) \epsilon 
\ff
 & = &
- {1 \over 2} K^2 \deps \mu - K \deps \sigma - \deps E
\, .
\eae
%%%%%%%%%%%%%%%%%%%%%%%%%%%%%%%%%%%%%%%%%%%%%%%%%%%%%%%%%%%%%%%%
\subsection{\mathversion{bold}%
The $\sigma$-identities
\mathversion{normal}}
%%%%%%%%%%%%%%%%%%%%%%%%%%%%%%%%%%%%%%%%%%%%%%%%%%%%%%%%%%%%%%%%
We introduce a number of functions on the loop,
relevant in the study of the solutions of~(\ref{epsde4}).
Any vector field $\vec a$ defines the following
two functions on the loop
\be
\label{anatdef}
a_n \equiv \vec a \cdot \nh
\, , \qquad \qquad
a_t \equiv \vec a \cdot \th
\, .
\ee
The position vector $\vec X$, in particular, supplies
\be
\label{phdef}
h \equiv \vec X \cdot \nh
\, , \qquad \qquad
p \equiv \vec X \cdot \th
\, ,
\ee
both functions evidently dependent on the choice of origin.
One finds
\be
\label{anathpid}
a_n'=-Ka_t
\, , \qquad \qquad
a_t'=Ka_n
\, , \qquad \qquad
h'=Kp
\, , \qquad \qquad
p'=1-Kh
\, ,
\ee
where $\vec a$ is henceforth assumed constant.
We will say that a function $f$ on the loop is the {\em generator}
of a transformation iff, under the latter, $\delta \vec X
\cdot \nh = f$. With this definition, 
$a_n$, $a_t$ are the generators of translations along 
$\vec a$ and perpendicularly to it respectively. $p$ and
$h$ generate rotations and dilations respectively (both
w.r.t.{} the origin). One easily verifies that
$DE_4(a_n)=DE_4(a_t)=DE_4(p)=0$
while $DE_4(h)=2\mu K + 3\sigma$, in accordance with the
scaling behavior of $\mu$, $\sigma$ found earlier,
Eq.~(\ref{eq:Pscale}).  
Experimenting a little with~(\ref{epsde4}) we find, not
without some surprise, two more solutions
\be
\label{twomore}
DE_4(K')=0
\, , \qquad \qquad
DE_4 \left(\sigma^{-1} (K^2 -\mu)\right) = 2 \mu K + 3\sigma
\, .
\ee
Noting, additionally, that when the origin is at the center of the
loop, the zeros of $p$ coincide with the extrema of $K$,
as well as the coincidence of the extrema of $h$ and
$K^2$, we make the ansatz
\be
\label{hpans}
h = \sigma^{-1}(K^2 -\mu) +f
\, , \qquad \qquad
p=2 \sigma^{-1} K' + g
\, ,
\ee
with $f$, $g$ to be determined. Substituting
in~(\ref{anathpid}), we find $f'=Kg$, $g'=-Kf$, with
solution
\be
f = a \cos(\Theta -\phi_0)
\, , \qquad \qquad
g = a \sin(\Theta - \phi_0)
\, ,
\ee
where $a$, $\phi_0$ are arbitrary constants. 
One recognizes $f$, $g$ to be, respectively, the functions $a_n$,
$a_t$ corresponding to the position vector $\vec a$ of the center
of the loop, then~(\ref{hpans}) state
\fle{magicids}{%
h = \sigma^{-1} (K^2 - \mu) + a_n
\, , \qquad \qquad
p=2\sigma^{-1} K' +a_t
\, ,
}%
a remarkable geometrical property, the implications of which will
occupy us shortly. Notice that neither of~(\ref{magicids})
survives in the $\sigma \rightarrow 0$ limit. Unless otherwise stated, 
we will take $\vec a$ equal to zero in what follows.

One might wonder whether~(\ref{epsde4}) admits other polynomial
solutions in $K$, apart from the second of~(\ref{twomore}). To
investigate this, we rewrite~(\ref{epsde4}) in terms of
derivatives w.r.t. $K$, using $\frac{d}{d\ell} = K' \frac{d}{d
K}$ and find for $DE(K^n)$ the leading term $(\frac{1}{8} n^4 +
\frac{3}{4} n^3 + \frac{1}{8} n^2 -3n -\frac{5}{2}) K^{n+4}$,
with only positive integer root $n=2$. We conclude that no other
polynomial solutions of~(\ref{epsde4}) exist.
%%%%%%%%%%%%%%%%%%%%%%%%%%%%%%%%%%%%%%%%%%%%%%%%%%%%%%%%%%%%%%%%
\subsection{Geometrics}
%%%%%%%%%%%%%%%%%%%%%%%%%%%%%%%%%%%%%%%%%%%%%%%%%%%%%%%%%%%%%%%%
We are now
in a position to give a purely geometrical construction
of the equilibria. Indeed, starting from $X^2 = h^2 + p^2$ and
using~(\ref{magicids}), we
find
\fle{geomproof}{%
X^2 -X_0^2  = 4 \sigma^{-1} K
\, , \qquad \qquad X_0 \equiv \sigma^{-1}\sqrt{8E+\mu^2}
\, ,
}%
where, in the derivation, use was made of the first 
integral, Eq.~(\ref{eq:D2}). This remarkable formula
expresses the embedding completely in terms of
$K(\ell)$. We emphasize that the shape is obtained
directly from $K$, given $(\sigma, \mu, E)$, without
any integration.
%The model is completely integrable.
Note that both the relations (\ref{magicids}) follow
by taking derivatives with respect to arclength of
(\ref{geomproof}). It follows from~(\ref{geomproof}) that the
moment of inertia $\calI$ of the loop around an axis
perpendicular to its plane and passing through its center, is
given by 
\begin{equation}
\calI = L X_0^2 +8\pi\sigma^{-1} 
\,.
\label{eq:K8}
\end{equation}

\par
%%%%%%%%%%%%%%%%%% FIGURE
\setlength{\figurewidth}{.9\textwidth}
\begin{figure}
%\setlength{\figurewidth}{.48\textwidth}
%\begin{floatingfigure}{.93\figurewidth}
%\begin{figure}
%\rule{0mm}{.675\figurewidth}
\begin{pspicture}(0\figurewidth,0\figurewidth)%
                 (.9\figurewidth,.63\figurewidth)
\setlength{\unitlength}{.25\figurewidth}
\psset{xunit=.25\figurewidth,yunit=.25\figurewidth,arrowsize=1.5pt 3}
%\psgrid[subgriddiv=10,griddots=5,gridlabels=8pt]
\psdots[dotscale=1.3]%
(2.807,.81)(1.9,.85)(2.31,1.06)
\put(2.83,.7){\makebox[0cm][l]{$P$}}
\put(2.32,.94){\makebox[0cm][r]{$A$}}
\put(2.07,.97){\makebox[0cm][r]{$X_0$}}
\put(2.18,.73){\makebox[0cm][r]{$X$}}
\put(3.42,1.34){\makebox[0cm][l]{$\rho=\frac{1}{K}$}}
\psline[linewidth=.2mm]{->}%
(1.9,.85)(2.31,1.06)
\psline[linewidth=.2mm]{->}%
(1.9,.85)(2.801,.81)
\psline[linewidth=.2mm]{<->}%
(3.37,1.055)(3.37,1.675)
%\put(1.1,1){\makebox[0cm][l]{$\cal{O}$}}
%\put(.85,1.25){\makebox[0cm][l]{$\vec a$}}
%%%%%%% Included graphic
\centerline{%
\raisebox{\totalheight}{%
\includegraphics[angle=270,width=.83\figurewidth]{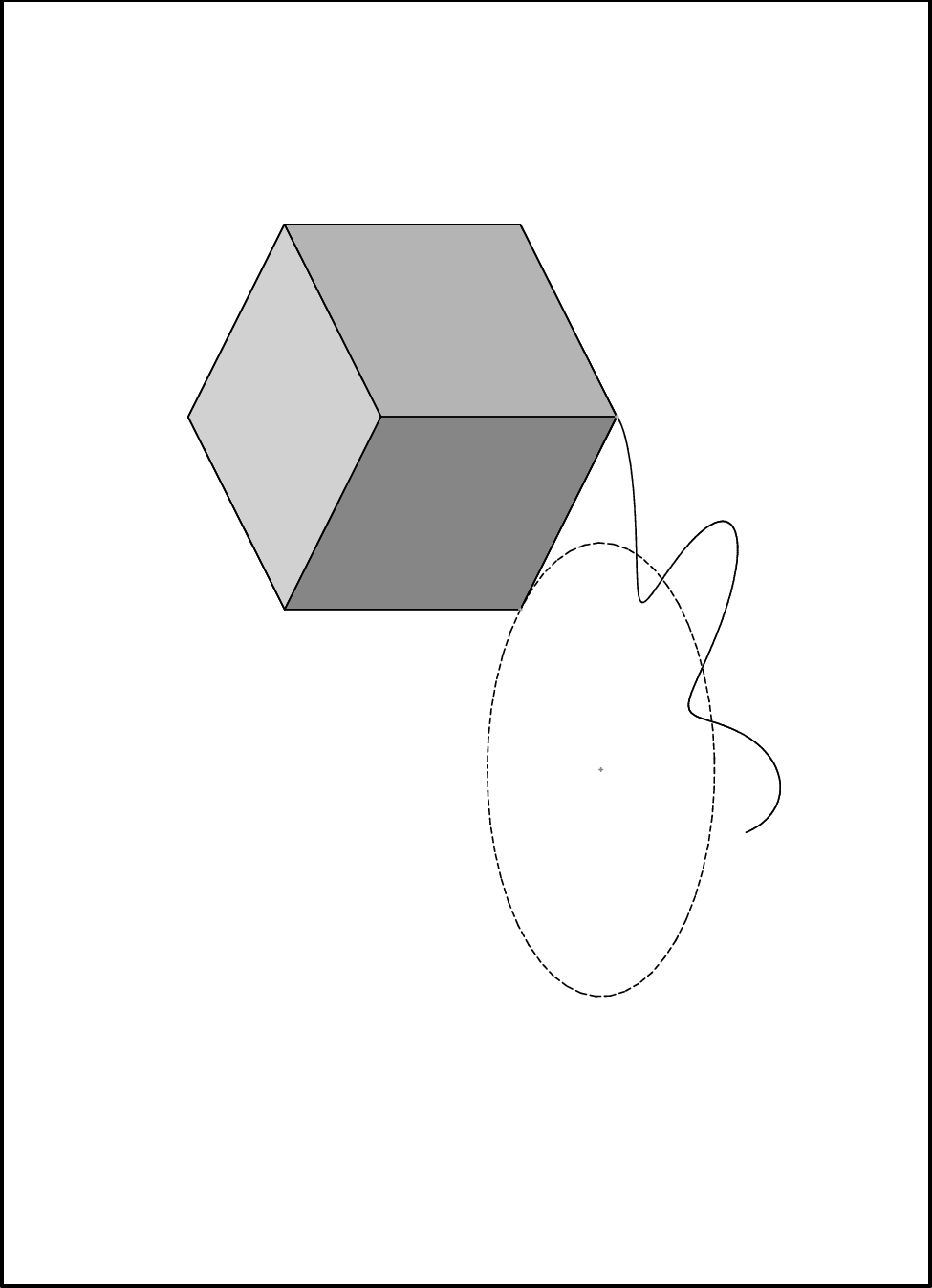}%
}%
}
\rput(-2.1,2){\rnode{A}{$\mbox{Volume}=4 \sigma^{-1}$}}
\rput(-1.8,1.7){\rnode{B}{}}
\nccurve[angleA=-45,angleB=100,linewidth=.18mm]{->}{A}{B}
%\put(-2,2){\makebox[0cm][l]{$\mbox{Volume}=4 \sigma^{-1}$}}
\end{pspicture}%
%}%
\capitem{\textsf{Geometrical construction of the equilibrium
curve. The parallelepiped has square base with side $AP$ and
volume $4\sigma^{-1}$. Its height $\rho$ gives the radius of 
curvature at $P$. The part
of the curve lying in the interior of the circle is constructed
similarly.}}
\label{construct}
%\end{floatingfigure}
\end{figure}
%%%%%%%%%%%%%%%%%% FIGURE
\par

\shortpage
\shortpage

To construct geometrically the equilibrium curve (for given
$\sigma$, $\mu$, $E$), we draw a circle with
radius $X_0$ and from a point $P$ outside it, bring the
tangent $PA$ to the circle (see Fig.~\ref{construct}). Taking as
base the square of this tangent, we construct a
parallellepiped with volume $4\sigma^{-1}$, the
height of which is the radius of curvature of the
equilibrium curve at $P$ --- this follows from~(\ref{geomproof}), 
written in the form $(X^2-X_0^2)\rho=4\sigma^{-1}$, where $\rho$
is the radius of curvature at $P$. To achieve closure of the
resulting curve, one has to start with a particular
slope, given by either of~(\ref{magicids}). 
In this way, one constructs the part of the curve lying
outside the circle (the latter intersects the
curve at its inflection points, if any) --- the interior part, 
present only if the ficticious particle reaches into the negative $K$ 
region, is constructed in a similar manner.  

\shortpage
\shortpage

Referring back to our expression for the force $\vec T(\ell)$,
Eq.~(\ref{summtf}), we realize that~(\ref{magicids}) implies that $\vec
T(\ell)$ is orthogonal to $\vec X(\ell)$ while $T(\ell) = \sigma
X(\ell)$, \ie,
\fle{TXrel}{%
\vec T(\ell) = \sigma ( \nh \times \th \, ) \times \vec X(\ell)
\, , \qquad \qquad
T_n = -\sigma p
\, , \qquad \qquad
T_t = \sigma h
\, ,
}%
with $\vec T(\ell)$ the force on the part of the loop pointed to
by $\th$. 
Notice that $\vec T$ is generally compressing but changes
nevertheless to true
tension at the points were $\vec X$ is tangential to the loop.
Also, $p$ has acquired a direct physical interpretation as a
consequence of~(\ref{TXrel}): $\sigma p d\ell$ is just the torque, 
w.r.t.{} the origin,
of the force due to the pressure on a segment of the loop with
length $d\ell$. Then the vanishing of the total torque on the
loop is guaranteed by the fact that $p$ is a derivative. In fact,
one may derive a compact and rather pleasing formula for the torque 
due to pressure
on any segment of the loop, like the one defined by $A$, $B$ in
the sketch of an $n=4$ configuration of Fig.{}~\ref{tcompf}.

\par
%%%%%%%%%%%%%%%%%% FIGURE
\begin{floatingfigure}{.25\textwidth}
\framebox{%
\begin{pspicture}(0\textwidth,0cm)(.2\textwidth,.2\textwidth)
\setlength{\unitlength}{.1\textwidth}
\psset{xunit=.1\textwidth,yunit=.1\textwidth,arrowsize=1.5pt 3}
%\psgrid[subgriddiv=10,griddots=5,gridlabels=8pt]
\psccurve(0,1)(.6,1.4)(1,2)(1.4,1.4)(2,1)(1.4,.6)(1,0)(.6,.6)(0,1)
\psline[linewidth=.2mm]{->}%
(.6,1.4)(1,0)
\psline[linewidth=.2mm]{->}%
(.6,1.4)(1,1)
\psdots[dotscale=1.3]%
(.6,1.4)(1,0)(1,1)
\put(.5,1.4){\makebox[0cm][r]{$A$}}
\put(1.,.15){\makebox[0cm][l]{$B$}}
\put(1.1,1){\makebox[0cm][l]{$\cal{O}$}}
\put(.85,1.25){\makebox[0cm][l]{$\vec a$}}
\end{pspicture}%
}%
\capitem{\textsf{Computing the torque due to pressure on the arc
$AB$}}
\label{tcompf}
\end{floatingfigure}
%%%%%%%%%%%%%%%%%% FIGURE
\par

Suppose we take $A$ as reference point,
then~(\ref{magicids}) is valid with $\vec a$ connecting $A$ with
the center $\cal{O}$ of the loop. We get
\[
\tau^{(A)}_{AB} \, = \, \sigma \int_A^B d\ell p
\, = \,
\sigma \int_A^B d\ell (2\sigma^{-1} K' +a_t)
\, = \,
2(K_B -K_A) + \sigma \vec a \cdot \int_A^B \vec{d\ell}
\]
\ie,
\fle{torqAB}{%
\tau^{(A)}_{AB} = 2(K_B - K_A) + \sigma \vec a \cdot
\overrightarrow{AB}
\, .
}%
Moving the reference point simply moves one endpoint of $\vec a$.
For $\vec a=0$ (torque w.r.t.{} $\cal{O}$),
$\tau^{(O)}_{AB} = 2(K_B - K_A)$. 
%%%%%%%%%%%%%%%%%%%%%%%%%%%%%%%%%%%%%%%%%%%%%%%%%%%%%%%%%%%%%%%%
\subsection{Self-intersections}
\label{osi}
%%%%%%%%%%%%%%%%%%%%%%%%%%%%%%%%%%%%%%%%%%%%%%%%%%%%%%%%%%%%%%%%
Given that some potential applications of our model
exclude self-intersecting configurations, we look now for
sufficient conditions for non-self-intersection.
We restrict our attention to self-intersections
that can be reached by a continuous deformation of
non-self-intersecting configurations. In other words, we consider
a one-parameter family of configurations $\vec X (\ell, t)$, $t
\in [0,1]$, continuous in $t$, such that, for every $t$, the
corresponding curvature satisfies~(\ref{eq:D1}) and we take $\vec X
(\ell,0)$, $\vec X (\ell,1)$ to be non-self-intersecting and
self-intersecting respectively. Then, we observe that  as $t$
varies from 0 to 1, one necessarily encounters a ``kiss":
%%%%%%%%%%%%%%%%%%%
\raisebox{0ex}[1.2\height][.6\height]{%
\begin{pspicture}(0ex,1.5ex)(5ex,4.5ex)
\psset{xunit=.6ex,yunit=.5ex}
\setlength{\unitlength}{.6ex}
%\psgrid[subgriddiv=2,griddots=2,gridlabels=8pt]
\put(4.5,2.5){$A$}
\psline[linewidth=.15mm]{*-*}%
(3,3.5)(3,3.5)
\pscurve[linewidth=.2mm]{-}%
(1,0)(3,3.5)(1,7.5)(3,9)(5,7.5)(3,3.5)(5,0)
\end{pspicture}
}%
%%%%%%%%%%%%%%%%%%%
The position vector $\vec X_A$ is along the axis of symmetry of
the lobe and tangential to the loop at $A$. The third
of~(\ref{TXrel}) then shows that the force at $A$ is normal to
the loop, while the second of~(\ref{TXrel}) gives its magnitude as 
\ble{Tnkiss}
T_{A} = \sigma X_{A}
\, .
\ee
The tangency of $\vec X$ at $A$ implies that $h_{A}=0$ and
hence, using~(\ref{magicids}) once more, we get (assuming that
$K$ is negative at a kiss\footnote{%
This assumption is true for all configurations we have
studied numerically.})
\fle{Kkiss}{%
K_{A} = - \sqrt{\mu}
\, ,
}%
for all kisses, regardless of the order of the configuration. 
We conclude that {\em a sufficient condition for
non-self-intersection (of the type defined above), is
$\mu <0$}. (\ref{Kkiss}) gives 
${K'}^2_{A}$ in terms of $\sigma$, $\mu$, $E$
\ble{KpE}
{K'}^2_{A} = 2\left(E + \frac{1}{8} \mu^2 - \frac{1}{2}
\sigma \sqrt{\mu}\right)
\, .
\ee
On the other hand, from~(\ref{geomproof}) we find that
\ble{Xkiss}
X^2_A = \sigma^{-2} (8E + \mu^2) - 4 \sigma^{-1} \sqrt{\mu} 
\, .
\ee
The tangency of $X_A$ though means that $2\sigma^{-1} K'_A=p_A = X_A$
and the above two equations then give (setting $\sigma=1$)
\ble{Emukiss}
E^2 + {1 \over 4} \left( \mu^2 -4 \sqrt{\mu} - {1 \over 2}
\right) E + {1 \over 64} (\mu^2 -4 \sqrt{\mu})(\mu^2 -4
\sqrt{\mu} -1) =0
\, ,
\ee
with roots
\ble{Emuroots}
E_1 = {1 \over 2} \sqrt{\mu} - {1 \over 8} \mu^2 +{1 \over 8}
\, , \qquad \qquad
E_2 = {1 \over 2} \sqrt{\mu} - {1 \over 8} \mu^2
\, .
\ee
For the particular case $n=2$, $K'_{A}=0$ and we get
$E=E_2$. We have
seen in Sec.~\ref{Config} that the relation $\Theta_0 =2\pi/n$
defines a curve in the $(\mu, E)$-plane, consisting of all
parameter pairs giving rise to a configuration of order $n$. The
intersection of that curve with the ones just written above,
consists of the points 
$(\mu,E)$ giving rise to a kissing configuration of order $n$. 

%%%%%%%%%%%%%%%%%%% ASIDE
%\begin{aside}
Consider how~(\ref{TXrel}), (\ref{torqAB}) guarantee equilibrium in
some particular examples. First, look at an even-$n$
configuration, say, $n=2$:  
%%%%%%%%%%%%%%%%%%% PS PICTURE
\raisebox{0ex}[.4\height][.6\height]{%
\begin{pspicture}(-3ex,1.8ex)(13ex,5ex)
\psset{xunit=.8ex,yunit=.6ex}
\setlength{\unitlength}{.8ex}
%\psgrid[subgriddiv=2,griddots=2,gridlabels=8pt]
\put(-3,2.25){$A$}
\put(13,2.25){$B$}
\psccurve[linewidth=.2mm]%
(3,6)(6,4)(9,6)(12,3)(9,0)(6,2)(3,0)(0,3)
\psline[linewidth=.15mm]{*-*}%
(0,3)(12,3)
%\psline{*-*}%
%(6,3)(6,3)
\psdots[dotscale=.6](6,3)
\end{pspicture}
}%
%%%%%%%%%%%%%%%%%%% PS PICTURE
The total force due to the pressure, pushing together the two
halves in the sketch, is $\sigma | \vec{AB}|$. (\ref{TXrel}) 
says that
the tension at $A$, $B$ (purely tangential, compressing) is
$\sigma |\vec{AB}|/2$, thus leaving each half at rest. As a second
example, consider the lobe defined by a self-intersection: 
%%%%%%%%%%%%%%%%%%% FIGURE
\raisebox{0ex}[1.2\height][.6\height]{%
\begin{pspicture}(-.5ex,1.5ex)(5ex,4.5ex)
\psset{xunit=.6ex,yunit=.5ex}
\setlength{\unitlength}{.6ex}
%\psgrid[subgriddiv=2,griddots=2,gridlabels=8pt]
\put(4.5,2.5){$A$}
\psline[linewidth=.15mm]{*-*}%
(3,3.5)(3,3.5)
\pscurve[linewidth=.2mm]{-}%
(0,0)(3,3.5)(6,7.5)(3,9)(0,7.5)(3,3.5)(6,0)
\end{pspicture}
}%
%%%%%%%%%%%%%%%%%%% FIGURE
The total force due to the pressure on the lobe is zero,
and~(\ref{TXrel}) says that the forces from the rest of the loop, 
on the two ends of the lobe that meet at $A$, are opposite (with
direction so as to keep the lobe closed). 
As a further check on our results, one can
verify
the balancing of torques using~(\ref{torqAB}) on, say, the right half
of a kiss.
%\end{aside}
%%%%%%%%%%%%%%%%%%% ASIDE
%%%%%%%%%%%%%%%%%%%%%%%%%%%%%%%%%%%%%%%%%%%%%%%%%%%%%%%%%%%%%%%%
\subsection{Connecting equilibria}
\label{ce}
%%%%%%%%%%%%%%%%%%%%%%%%%%%%%%%%%%%%%%%%%%%%%%%%%%%%%%%%%%%%%%%%
There is  more to be derived from~(\ref{geomproof}).
Taking $\deps$ on both sides, we find
\be
\label{var1}
DE_2^{(a)}(\epsilon) 
\equiv 
2 \epsilon'' + (3 K^2 -\mu)\epsilon = \frac{\sigma}{2}  \deps
X_0^2 -2K \sigma^{-1} \deps \sigma
\, ,
\ee
a considerable improvement over~(\ref{epsde4})\footnote{Notice,
however, that~(\ref{var1}) is only valid when the center of
the loop is at the origin.}. Moreover,
by varying the first of~(\ref{magicids}) we also find   
\be
\label{var2}
DE_2^{(b)}(\epsilon) 
\equiv 
2 K \epsilon'' -2K'\epsilon' + (\sigma +2K^3)\epsilon =
-\sigma^{-1} (K^2-\mu) \deps \sigma -\deps \mu
\, .
\ee
Comparison with~(\ref{var1}) leads to a first order
equation for $\epsilon$
\be
\label{firstord}
DE_1(\epsilon) 
\equiv 
4K' \epsilon' -4 K'' \epsilon = c_2 K^2 + c_1 K + c_0
\, ,
\ee
where the constants $c_i$ are given by
\ble{cidef}
c_2 \equiv  -2 \sigma^{-1} \deps \sigma 
\, , \qquad \qquad
c_1 \equiv \sigma \deps X_0^2
\, , \qquad \qquad
c_0 \equiv 2 \deps \mu - 2\mu \sigma^{-1} \deps \sigma
\, .
\ee
Noting that ${d \over dK} =(K')^{-1} {d \over
d\ell}$,~(\ref{firstord}) can be written in the form 
\be
\label{first2}
8(E-V) \dot{\epsilon} +4\dot{V}\epsilon =c_2 K^2 + c_1 K
+ c_0
\, ,
\ee
where the dot denotes differentiation w.r.t. $K$ and
$V=V(K)$. This latter equation is readily integrated to
give
\fle{epsiatlast}{%
\dss
\epsilon = {1 \over 4} K' \int d K {c_2
K^2 + c_1 K + c_0 \over \big(2(E-V) \big)^{3/2}}
\, .
}%
One may add an arbitrary amount of $K'$ (rotation) to this (but
{\em not} $a_n$ or $a_t$, since these move the origin). 
To get particular solutions from~(\ref{epsiatlast}), we
need to specify the direction of the deformation in the
$(\sigma,\, \mu)$-plane, \ie, \ the ratio $\deps \mu
/\deps \sigma$. 
It will prove convenient for our further analysis 
of~(\ref{epsiatlast}), to reparameterize the $(\sigma,\, 
\mu)$-plane introducing new coordinates $(\lambda, \xi)$ via
\ble{smlk}
\lambda(\sigma, \mu) =\sigma^{-1/3}
\, , \qquad \qquad
\xi(\sigma, \mu) = \sigma^{-2/3} \mu
\, .
\ee

\par
%%%%%%%%%%%%%%%%%% FIGURE
\begin{floatingfigure}{.32\textwidth}
\vspace{2mm}
\framebox{%
\begin{pspicture}(-.04\textwidth,-.035\textwidth)%
(.22\textwidth,.19\textwidth)
\setlength{\unitlength}{.1\textwidth}
\psset{xunit=.1\textwidth,yunit=.1\textwidth,arrowsize=2pt 3}
%\psgrid[subgriddiv=10,griddots=5,gridlabels=8pt]
\psline[linewidth=.3mm]{->}%
(0,0)(2,0)
\psline[linewidth=.3mm]{->}%
(0,0)(0,1.6)
\pscurve(0,0)(.01,.0464)(.5,.629)(1,1)(1.25,1.16)(1.5,1.31)(2,1.587)
\pscurve{->}(0,0)(.01,.0464)(.5,.629)(1,1)(1.25,1.16)
\psline[linewidth=.2mm,linestyle=dashed]{-}%
(1,0)(1,1)
\psline[linewidth=.2mm,linestyle=dashed]{-}%
(0,1)(1,1)
\psline[linewidth=.2mm,linestyle=dashed]{-}%
(1.5,0)(1.5,1.31)
\psline[linewidth=.2mm,linestyle=dashed]{-}%
(0,1.31)(1.5,1.31)
\psdots[dotscale=1.3]%
(1,1)(1.5,1.31)
\put(1.05,.85){\makebox[0cm][l]{$P'$}}
\put(1.55,1.15){\makebox[0cm][l]{$P$}}
\put(.8,-.2){\makebox[0cm][l]{$\sigma \! = \! 1$}}
\put(1.45,-.2){\makebox[0cm][l]{$\sigma$}}
\put(-.1,1){\makebox[0cm][r]{$\xi$}}
\put(-.1,1.31){\makebox[0cm][r]{$\mu$}}
\put(1.1,1.12){\makebox[0cm][r]{$\lambda$}}
\end{pspicture}%
}%
\capitem{\textsf{Definition of the coordinates $\lambda$, $\xi$.
\\ }}
\label{laxi}
\end{floatingfigure}
%%%%%%%%%%%%%%%%%% FIGURE
\par

A point $P$ with coordinates $(\sigma,\mu) \neq (0,0)$ lies on a 
unique scaling orbit, which can be specified by the 
$\mu$-coordinate of its point of intersection $P'$ with the 
$\sigma=1$ 
line, this is the value of $\xi$ for $P$. One can get now from 
$P'$ to $P$ by scaling by $\lambda$ (see Fig.~\ref{laxi}). The
obvious advantage of these new coordinates is that the
scaling orbits are constant-$\xi$ lines. The dependence on
$\lambda$ of
an arbitrary quantity $S(\lambda,\xi)$, with length dimension
$q$, is $S(\lambda,\xi)=\lambda^q \tilde{S}(\xi)$, where we denote
by a tilde the remaining function of $\xi$. It follows that
${\partial S \over \partial \lambda} = q \lambda^{-1} S$ so that
$\deps S = q \lambda^{-1} S \deps \lambda + \lambda^{q} 
\tilde{S}' \deps \xi$, the prime denoting here differentiation
w.r.t.{} $\xi$.

%%%%%%%%%%%%%%%%%% ASIDE
%\begin{aside}
We verify that~(\ref{epsiatlast}) gives $\epsilon \sim
h$ for scaling. Eq.~(\ref{cidef}), written in terms of
$(\lambda,\xi)$, gives
\bae
c_2
\fe
6 \lambda^{-1} \deps \lambda
\ff
c_1
\fe 
\lambda^{-2} \left( 16 \tilde{E} +2  \xi^2
\right) \deps \lambda + \lambda^{-1} \left( 8 
\tilde{E}' + 2  \xi \right) \deps \xi
\\
c_0 
\fe
2 \lambda^{-3} \xi \deps \lambda + 2 \lambda^{-2} \deps \xi
\, ,
\nn
\label{cidef2}
\eae
where $E(\lambda,\xi) = \lambda^{-4} \tilde{E}(\xi)$ is the energy
that guarantees closure of some particular configuration. 
An increment $\deps \lambda$ corresponds to scaling by a
factor $1+ \deps \lambda / \lambda$, hence the
corresponding $\epsilon$ should be $\epsilon = (\deps \lambda
/ \lambda) h$. Putting $\deps \xi=0$ in~(\ref{cidef2})
one determines the $c_i$ for pure scaling, then computing $d
/ dK \bigl((K^2 - \mu) / K')$ one finds that the $K^5$ and
$K^3$ terms in the numerator cancel and one recovers the
integrand in~(\ref{epsiatlast}) with just the right $c$'s.
%\end{aside}
%%%%%%%%%%%%%%%%%% ASIDE

Integrating the first of~(\ref{magicids}) w.r.t. $d\ell$,
and using 
\be
\label{Ainth}
A = {1 \over 2} \int d\ell h
\, ,
\ee
we get
\be
\label{WAL}
F = \mu L + 2 \sigma A
\, ,
\ee
from which
\be
\label{Wvar1}
\deps F = L \deps \mu  + \mu \deps L + 2 A \deps \sigma  +
2 \sigma \deps A
\ee
follows.  On the other hand, direct variation of $F$,
Eq.~(\ref{Fzdef}), 
gives
\be
\label{Wvar2}
\deps F = - \mu \deps L - \sigma \deps A
\, .
\ee
Comparing with~(\ref{Wvar1}) we find 
\be
\label{comprel}
L \deps \mu + 2 A \deps \sigma  + 3 \sigma \deps A 
+ 2 \mu \deps L =0
\, .
\ee
For a length-preserving deformation, $\deps L=0$,
and~(\ref{comprel}) reduces to
\be
\label{musirat}
L \deps \mu + 2 A \deps \sigma  
+ 3 \sigma \deps A =0
\, .
\ee
We note in passing that the various differential operators we have 
defined above are related in the following way
\bae
\label{relDE}
[ DE_1 (\epsilon ) ] ' 
\fe 2 K' DE_2{}^{(a)}  (\epsilon )
\, ,
\ff
K [ DE_1  (\epsilon )] ' 
\fe 
2 K' DE_2{}^{(b)} (\epsilon ) + K' DE_1 (\epsilon ) 
\,,
\ff
DE_3 (\epsilon ) 
\fe 
K' [ DE_2{}^{(a)} (\epsilon ) ] ' - K''  DE_2{}^{(a)} (\epsilon ) 
- {1 \over 4} (K^2 - \mu ) DE_1 (\epsilon ) 
\, , 
\ff
K'  DE_4 (\epsilon )   
\fe 
2 [ DE_3 (\epsilon ) ]' 
\, . 
\nonumber
\eae
It follows that any perturbation that satisfies Eq.~(\ref{firstord})
will necessarily satisfy all the higher order ones. 
%%%%%%%%%%%%%%%%%%%%%%%%%%%%%%%%%%%%%%%%%%%%%%%%%%%%%%%%%%%%%%%%
\subsection{Perturbing the circle}
\label{ptc}
%%%%%%%%%%%%%%%%%%%%%%%%%%%%%%%%%%%%%%%%%%%%%%%%%%%%%%%%%%%%%%%%
Referring to our cylinder model of the loop (see Sec.~\ref{oft}), 
we consider here the following thought experiment: we start with
the loop ``filled'' to capacity, \ie, in a circular
configuration, and start deflating it by removing fluid from its
interior while keeping its length fixed. We would like to follow 
the evolution of its shape, for
distinct $n$'s, and derive the limiting form of the bending
energy $F$ as a function of the area $A$ near the circular
extreme. Going beyond the cylinder model, we would also like to
allow for self-intersections --- what happens if one just keeps
subtracting area?

Consider the following perturbation to a circle of radius $R$
\ble{pertcirc}
\rho(\theta) = R_{\alpha}\bigl(1+\alpha \sin(n \theta)\bigr)
\, , 
\qquad \qquad
\alpha \ll 1
\, .
\ee
The line element and curvature in these coordinates are 
\ble{lecurv}
d \ell = \sqrt{\rho^2 + \dot{\rho}^2} d\theta
\, ,
\qquad \qquad
K = \frac{\rho^2 +2 \dot{\rho}^2 -\rho
\ddot{\rho}}{(\rho^2+\dot{\rho}^2)^{3/2}}
\, ,
\ee
where the dot here denotes differentiation w.r.t. $\theta$.
Substituting~(\ref{pertcirc}) in the first of~(\ref{lecurv}), 
integrating over $\theta$ and requiring the total length to be
equal to the initial value $2\pi R$, we find
\ble{Ralpha}
R_\alpha = R(1 - \frac{n^2 \alpha^2}{4}) + \calO (\alpha^3)
\, ,
\ee
while for the curvature the second of~(\ref{lecurv}) gives
\ble{curvpert}
K = \frac{1}{R} \bigl(1 + \alpha (n^2-1) \sin(n \theta)\bigr) +
\calO(\alpha^2)
\, .
\ee
We substitute the above expression in the differential equation
for $K$, Eq.~(\ref{eq:D1}), and demand that it be a solution to  
$\calO(\alpha)$. The
constant and $\calO(\alpha)$ terms respectively give
\ble{a01}
1-R^2 \mu - R^3 \sigma=0 
\, ,
\qquad \qquad
2n^4 + (R^2 \mu -5) n^2 + 3 - R^2 \mu =0
\, .
\ee
Notice that the first relation is simply the statement
$\dot{V}(1/R)=0$ while the second also follows 
from~(\ref{epsde4}) with $\epsilon$ as in~(\ref{pertcirc}).
For this perturbation  the differential equation
cannot be satisfied to $\calO(\alpha^2)$. From~(\ref{a01}) 
we get
\ble{musipert}
\sigma = \frac{2}{R^3} (n^2-1)
\, ,
\qquad \qquad
\mu = \frac{1}{R^2} (3-2n^2)
\, .
\ee
During the deflating process, both $\sigma$ and $\mu$ will
vary as functions of $A$. The point in the $\sigma$-$\mu$ plane
corresponding to the configuration will trace out an orbit, 
starting at the above points (for each $n$), all of which lie on
the line $R^3 \sigma + R^2 \mu =1$. For $F$ and $A$ we find
\ble{FApert}
F = \frac{2\pi}{R} \bigl(1 + \frac{1}{2}(n^2-1)^2 \alpha^2
\bigr)
+\calO(\alpha^3)
\, ,
\qquad \qquad
A = \pi R^2 \bigl(1 - \frac{1}{2} (n^2-1) \alpha^2\bigr) 
     + \calO(\alpha^3)
\, .
\ee
Notice that, near the circle, 
\[ \frac{\delta F}{\delta
A}=-\frac{2}{R^3} (n^2-1) = - \sigma
\, ,
\]
in agreement with~(\ref{Wvar2}) (for $\delta L =0$). As 
$A$ keeps diminishing, our numerical analysis shows that the
configurations start self-intersecting, giving rise to regions of
negative area --- this scenario is sketched in Fig.{}~\ref{seque}.
Notice how the winding numbers ($+1$ for the little circles, $-1$ for
the big one) add up correctly to match that of the circle to
which this configuration can be continuously deformed. This
observation points to a feature of the above $m=1$ limiting
configurations already alluded to in the
introduction: the big circle in the configuration of order $n$
actually winds around itself $n-1$ times. For example, to move
from one lobe to the next  along the loop, in the $n=3$
configuration, one must travel an angle of $4\pi/3$, {\em not}
$2\pi/3$ (a glance at Fig.~\ref{configint}.b reveals how this
comes about). 
The limiting shape in this direction then is a circular one, 
with radius 
slightly less than $R/(n-1)$ and $n$ little circles attached to it. 
In this extreme, $F(A)$ is dominated by the little circles
and assumes the limiting form
\ble{FAm1}
F(A) \cong \frac{4\pi n^2}{R} \frac{1}{1+(n-1) \frac{A}{\pi R^2}}
\, ,
\ee
where the area is counted with the appropriate multiplicity due
to the winding (\eg, plus three times the area of the little
circle, minus twice that of the big one, in the $n=3$ case). 
A sketch of $F(A)$ that interpolates between~(\ref{FApert})
and~(\ref{FAm1}) is given in Fig.{}~\ref{FAplot}. 
%%%%%%%%%%%%%%%%%% FIGURE
\setlength{\figurewidth}{.9\textwidth}
\begin{figure}
\begin{center}
\begin{pspicture}(0\figurewidth,0\figurewidth)%
                 (\figurewidth,.25\figurewidth)
\setlength{\unitlength}{.25\figurewidth}
\psset{xunit=.25\figurewidth,yunit=.25\figurewidth,arrowsize=1.5pt 3}
%\psgrid[subgriddiv=10,griddots=5,gridlabels=8pt]
%%%%%%% Puts
%\put(2.6,.3){\makebox[0cm][l]{$X$}}
%\put(1.52,2.4){\makebox[0cm][r]{$Y$}}
%%%%%%% Axes
%\psline[linewidth=.2mm]{->}%
%(1,.45)(2.6,.45)
%\psline[linewidth=.2mm]{->}%
%(1.64,.45)(1.64,2.5)
%%%%%%% Curves
\pscircle%
(.5,.5){.4}
\psline[linewidth=.3mm]{->}%
(.97,.5)(1.15,.5)
\psccurve%
(1.5,.95)(1.8,.75)(1.7,.5)(1.8,.25)(1.5,.05)(1.2,.25)(1.3,.5)(1.2,.75)
\psline[linewidth=.3mm]{->}%
(1.9,.5)(2.08,.5)
\psccurve%
(2.5,.92)(2.65,.85)(2.2,.5)(2.65,.15)(2.5,.08)%
(2.35,.15)(2.8,.5)(2.35,.85)
\psline[linewidth=.3mm]{->}%
(2.87,.5)(3.05,.5)
\pscircle%
(3.5,.5){.38}
\pscircle%
(3.5,.915){.04}
\pscircle%
(3.5,.085){.04}
%\psccurve%
%(3.5,.94)(3.53,.91)(3.5,.88)(3.12,.5)(3.5,.12)(3.53,.09)(3.5,.06)%
%(3.47,.09)(3.5,.12)(3.88,.5)(3.5,.88)(3.47,.91)
\end{pspicture}%
\end{center}
\capitem{Evolution of a $n=2$ configuration under deflation
(sketch). The central region in the third configuration, as well
as the big circle in the fourth one, contribute negative area.}
\label{seque}
\end{figure}
%%%%%%%%%%%%%%%%%% FIGURE

\par
%%%%%%%%%%%%%%%%%% FIGURE
\setlength{\figurewidth}{.7\textwidth}
%\begin{floatingfigure}{.93\figurewidth}
\begin{figure}
%\rule{0mm}{.675\figurewidth}
\begin{pspicture}(0\figurewidth,0\figurewidth)%
                 (.9\figurewidth,.7\figurewidth)
\setlength{\unitlength}{.25\figurewidth}
\psset{xunit=.25\figurewidth,yunit=.25\figurewidth,arrowsize=1.5pt 3}
%\psgrid[subgriddiv=10,griddots=5,gridlabels=8pt]
%%%%%%% Puts
\put(2.87,2.42){\makebox[0cm][r]{$\bar{F}$}}
\put(4.35,.149){\makebox[0cm][l]{$\frac{A}{\pi R^2}$}}
\put(1.15,2.38){\makebox[0cm][l]{$n=$}}
\put(1.45,2.38){\makebox[0cm][l]{$2$}}
\put(2.2,2.38){\makebox[0cm][l]{$3$}}
\put(2.46,2.38){\makebox[0cm][l]{$4$}}
\put(2.61,2.38){\makebox[0cm][l]{$5$}}
%%%%%%% Axes
\psline[linewidth=.2mm]{->}%
(2.83,.27)(2.83,2.4)
\psline[linewidth=.2mm]{->}%
(1.41,.265)(4.4,.265)
%%%%%%% Included graphic
\centerline{%
\raisebox{\totalheight}{%
\includegraphics[angle=270,width=.9\figurewidth]{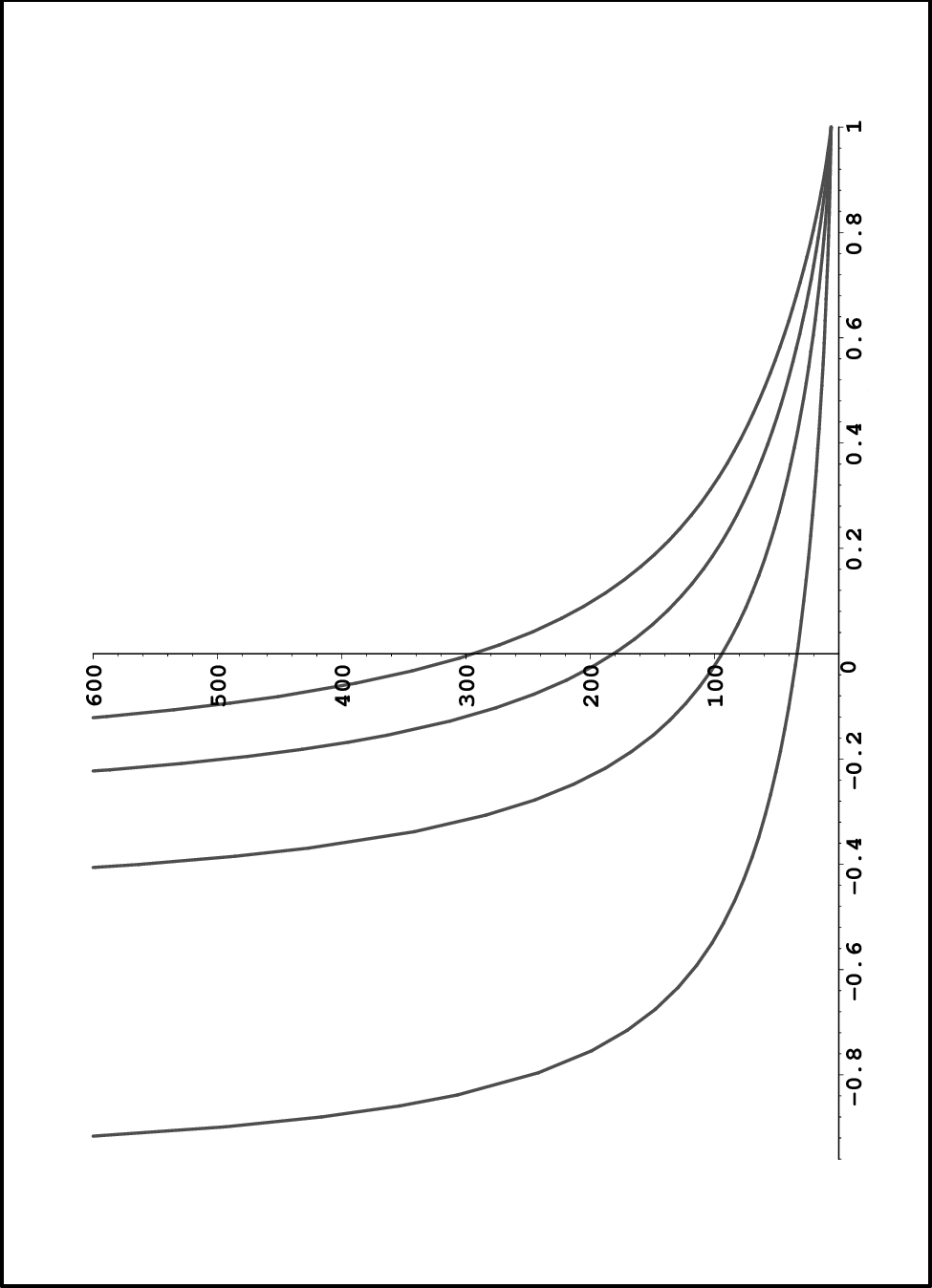}%
}%
}
\end{pspicture}%
\capitem{Bending energy $\bar{F}$ {\em vs.} area $A/\pi R^2$ for
$n=2,3,4,5$ (interpolation). All curves touch at the point on the
right, which corresponds to a circular configuration, giving rise
to bifurcation. The vertical asymptotes are at $-\frac{1}{n-1}$.}
\label{FAplot}
\end{figure}
%\end{floatingfigure}
%%%%%%%%%%%%%%%%%% FIGURE
\par

%%%%%%%%%%%%%%%%%%%%%%%%%%%%%%%%%%%%%%%%%%%%%%%%%%%%%%%%%%%%%%%%
\section{Conclusions}
\label{Conc}
%%%%%%%%%%%%%%%%%%%%%%%%%%%%%%%%%%%%%%%%%%%%%%%%%%%%%%%%%%%%%%%%
In this paper, we have studied the equilibrium configurations
of elastic planar loops  with constant area and constant
length. 
The condition of closure of the
loop gives a discrete spectrum of configurations, lying along
several branches in parameter space.
We focus on non self-intersecting loops that
inflate to a circle when the enclosed area is increased, due to
their relevance as toy model 
for two-dimensional membranes.  For this branch,
starting from analytical expressions
in the relevant limit cases, we obtain a reliable
sketch of the dependence of the energy
on the area. The equation that determines equilibria
connecting perturbations led us, rather unexpectedly, to the
$\sigma$-identities~(\ref{magicids}), from which
a novel geometrical construction of the (constrained) elastic
loop followed.
In turn, these identities permit the expression 
of an equilibrium connecting normal deformation in terms
of the curvature and the appropriate perturbations of
the parameters.

In future work, we plan to address the important issue
of the stability of the equilibrium configurations.
In light of the complicated structure of the fourth
order differential operator appearing in (\ref{epsde4}), this
appears to be a non-trivial task in the general case.
It is indicative of the intrinsic complexity of the question that
even the relatively simple case of the figure eight configuration
requires the elaborate analysis of~\cite{Ive.Sin:97}.

Another interesting issue is the analysis of thermal fluctuations 
in this model, which would provide an analytical
counterpart to the study of Ref.~\cite{Fis.Lei.Sin:87} of 
two-dimensional
vesicles, using Monte Carlo techniques. In particular,
there is a similarity between
some of the configurations of Fig.~2 of the above work and the
configuration of Fig.~\ref{config}.a in this paper.
It is also interesting to note
that the shapes of Fig.~\ref{config} closely resemble a top view
of the starfish vesicles of Ref.~\cite{Win.Dob.Sei:96},
which are almost planar. Finally, configurations similar to ours
appear in~\cite{Tad.Ode:67}, where our problem has been
approached from a functional analytic point of view and important
existence results have been derived, as well as in~\cite{Fuk:97}. 

After the completion of this paper, we set out to study its
most natural extension: geometric models for loops in space.
We became aware of a large body of literature that explores 
the interconnections among hierarchies of functionals of the
geometry of a curve, 
their corresponding
generators of rigid motion and integrable systems, such as
the KdV equation. We have found Ref.~\cite{Lan:99} an excellent point
of entry to the subject.
We also realized that our $\sigma$-identities can be obtained by
adapting the cylindrical coordinates used by Langer and
Singer~\cite{Lan.Sin:84a} in the analysis of buckled rings.
We expect that these explorations, apart from their intrinsic
interest,  will contribute to a deeper
understanding of two-dimensional membranes.
%%%%%%%%%%%%%%%%%%%%%%%%%%%%%%%%%%%%%%%%%%%%%%%%%%%%%%%%%%%%%%%%%
%%%%%%%%%%%%%%%%%%%%%%%%%%%%%%%%%%%%%%%%%%%%%%%%%%%%%%%%%%%%%%%%%
%%%%%%%%%%%%%%%%%%%%%%%%%%%%%%%%%%%%%%%%%%%%%%%%%%%%%%%%%%%%%%%%%
%%%%%%%%%%%%%%%%%%%%%%%%%%%%%%%%%%%%%%%%%%%%%%%%%%%%%%%%%%%%%%%%%
%%%%%%%%%%%%%%%%%%%%%%%%%%%%%%%%%%%%%%%%%%%%%%%%%%%%%%%%%%%%%%%%%
%%%%%%%%%%%%%%%%%%%%%%%%%%%%%%%%%%%%%%%%%%%%%%%%%%%%%%%%%%%%%%%%%
\section*{Acknowledgements}
G.A. wishes to thank Gilberto Tavares for technical assistance
and CONACyT for a graduate fellowship. R.C. was supported by
CONACyT grant 32187-E. G.A., C.C. and J.G. were supported by
CONACyT grant 32307-E and DGAPA-UNAM grant IN119792. 
%%%%%%%%%%%%%%%%%%%%%%%%%%%%%%%%%%%%%%%%%%%%%%%%%%%%%%%%%%%%%%%%
%%%%%%%%%%%%%%%%%%%%%%%%%%%%%%%%%%%%%%%%%%%%%%%%%%%%%%%%%%%%%%%%%
\appendix
%%%%%%%%%%%%%%%%%%%%%%%%%%%%%%%%%%%%%%%%%%%%%%%%%%%%%%%%%%%%%%%%%
%%%%%%%%%%%%%%%%%%%%%%%%%%%%%%%%%%%%%%%%%%%%%%%%%%%%%%%%%%%%%%%%%
%\renewcommand{\thesection}{}
\appendix
\section{Some Useful Formulae}
\label{Dereom}
%%%%%%%%%%%%%%%%%%%%%%%%%%%%%%%%%%%%%%%%%%%%%%%%%%%%%%%%%%%%%%%%
%%%%%%%%%%%%%%%%%%%%%%%%%%%%%%%%%%%%%%%%%%%%%%%%%%%%%%%%%%%%%%%%%
Under an infinitesimal 
deformation of the loop along the normal, $\vec X
\rightarrow \vec X + \delta_\epsilon \vec X \equiv \vec X +
\epsilon(\ell) \nh$, we find
\[
\begin{array}{rclcrclcrcl}
\dss \deps d\ell \fedss K \epsilon d \ell
\, , \aqqadss
\deps \th \fedss \epsilon' \nh
\, , \aqqadss
\deps K \fedss -\epsilon'' - K^2 \epsilon 
\ff
\dss \deps \derl \fedss -K \epsilon \derl
\, , \aqqadss 
\deps \nh \fedss -\epsilon' \th 
\, ,  \aqqadss 
\deps K' \fedss 
-\epsilon''' -K^2 \epsilon' -3KK' \epsilon
\\[1mm] \nonumber
\deps h \fedss 
\epsilon -p\epsilon'
\, , \aqqadss 
\deps p \fedss
\epsilon' h
\, , \aqqadss 
\deps K'' \fedss
-\epsilon''''-K^2 \epsilon'' -5KK'\epsilon' \\
 & & & & & & & & & & \dss -(6E-{11 \over 4} K^4 + {7 \over 2}
\mu K^2 + 5 \sigma K)\epsilon
\end{array}
\]
(we have used the e.o.m and the first integral,
Eqs.~(\ref{eq:D1}) and (\ref{eq:D2}), to express
$K''$ and $(K')^2$ in terms of $K$). 

\noindent For the derivatives of $h$, $p$, $K$ we find
\bae
h' \fe Kp
\ff
h'' \fe -K^2 h + K' p +k
\ff
h''' \fe -3KK' h + (-{3 \over 2} K^3 +{\mu \over 2} K + {
\sigma \over 2}) p + 2K'
\ff
h'''' \fe \big(-6E +{15 \over 4} K^4 - {7 \over 2} \mu K^2 -5
\sigma K\big) h +\big(-{15 \over 2} K^2 K' + {\sigma \over 2}
K'\big)p -{5 \over 2} K^3 + {3 \over 2} \mu K + {3 \over 2}
\sigma 
\ff
 & & \ff
p' \fe 1 -Kh
\ff
p'' \fe -K' h -K^2 p
\ff
p''' \fe ({3 \over 2} K^3 -{\mu \over 2} K - {\sigma
\over 2})h -3KK' p -K^2
\ff
p'''' \fe ({15 \over 2} K^2 K' - {\mu \over 2} K') h 
+ ({15 \over 4} K^4 - {7 \over 2} \mu K^2 -5 \sigma K
-6E) p -5KK'
\ff
\ & & \ff
K' \fe \sqrt{2}(E-\frac{1}{8} K^4 +\frac{\mu}{4} K^2 +
\frac{\sigma}{2} K)^{1/2}
\ff
K'' \fe \frac{1}{2} (-K^3 +\mu K + \sigma)
\ff
K''' \fe -{3 \over 2} K^2 K' + {\mu \over 2} K'
\ff
K'''' \fe {3 \over 2} K^5 - {10 \over 4} \mu  K^3 - {15 \over
4} \sigma K^2 + ({\mu^2 \over 4} -6E)K + {\mu \sigma
\over 4}
\nonumber
\eae
%%%%%%%%%%%%%%%%%%%%%%%%%%%%%%%%%%%%%%%%%%%%%%%%%%%%%%%%%%%%%%%%%
\section{Averages}
\label{Ave}
%%%%%%%%%%%%%%%%%%%%%%%%%%%%%%%%%%%%%%%%%%%%%%%%%%%%%%%%%%%%%%%%%
Consider the quantities $W_n$, defined by
\ble{Wndef}
W_n = \int_0^{L_0} K^n d\ell  = 
2\int_{K_{min}}^{K_{max}} K^n \frac{dK}{K'} 
\, .
\ee
$L_0$ here is the length along the loop corresponding to one full
oscillation of the particle (we will use $F_0$ later with a
similar meaning). $W_n$ is then (proportional to) the average of
$K^n$ along the loop. Starting from 
\ble{intze}
\int_{K_{min}}^{K_{max}} \frac{d}{dK} (K^n K') =0
\, ,
\ee
one may derive the recursion relation~\cite{Abr.Ste:65}
\ble{recrel}
-\frac{1}{8} (n+2) W_{n+3} + \frac{\mu}{4} (n+1) W_{n+1} +
\frac{\sigma}{4} (2n+1) W_n + En W_{n-1}=0
\, ,
\qquad
n=0,1,2,\dots
\, ,
\ee
which permits, in principle, the calculation of the average of any 
power series in $K$, for a given equilibrium configuration. 
In particular,
\ble{W4W3}
W_3 = \mu W_1 + \sigma W_0
\, ,
\qquad 
\qquad 
\qquad 
W_4 = \frac{4}{3} W_2 +2\sigma W_1 +\frac{8}{3} E W_0
\, .
\ee

Referring back to Fig.~\ref{phase}, we notice that the area $S$
enclosed by the orbit of the phase point, for some given $\sigma$, 
$\mu$, $E$, is given by 
\ble{SmusE}
S(\sigma, \mu, E)=2\int_{K_{min}}^{K_{max}} K'dK
\, .
\ee
Differentiating w.r.t.{} $\sigma$ and using the fact that the
integrand vanishes at the endpoints,  we find
\ble{ThetaS}
\Theta_{0}= 2 \frac{\partial S}{\partial \sigma}
\, .
\ee
Length and bending energy per full oscillation 
also follow by simple differentiation
\ble{LFS}
L_0 =  \frac{\partial S}{\partial E}
\, , 
\qquad 
\qquad 
\qquad
F_0 = 4 \frac{\partial S}{\partial \mu}
\, .
\ee
Notice that these quantities are defined whether or not the
closure condition is satisfied, \ie, they are functions of the
three independent variables $\sigma$, $\mu$, $E$.
Combining~(\ref{SmusE}) and~(\ref{W4W3}) above, we get
\ble{Sexp}
S=\frac{4}{3} EL + \frac{1}{6} \mu F +\frac{1}{2} \sigma \Theta
\, .
\ee
One may regard $L_0$, $\Theta_0$ and $F_0$ as new coordinates in
the space of (not necessarily closed) configurations --- the
change of coordinates is non-singular, except for some special
points, and is given by a Legendre transform. Then 
the closure condition, Eq.~(\ref{condclos}),
restricts to the configurations that lie on the $L_0$-$F_0$
plane, with $\Theta_0=2\pi/n$. 
%%%%%%%%%%%%%%%%%%%%%%%%%%%%%%%%%%%%%%%%%%%%%%%%%%%%%%%%%%%%%%%%%
%%%%%%%%%%%%%%%%%%%%%%%%%%%%%%%%%%%%%%%%%%%%%%%%%%%%%%%%%%%%%%%%%
%%%%%%%%%%%%%%%%%%%%%%%%%%%%%%%%%%%%%%%%%%%%%%%%%%%%%%%%%%%%%%%%%
%%%%%%%%%%%%%%%%%%%%%%%%%%%%%%%%%%%%%%%%%%%%%%%%%%%%%%%%%%%%%%%%%
%\bibliographystyle{plain}
%\bibliography{strings}

\end{document}